\documentstyle[preprint,tighten,aps,prc,epsf,rotate]{revtex}
\begin{document}
\baselineskip 16pt plus 2pt minus 2pt
\author{Nimai C. Mukhopadhyay$^{a}$
\thanks{e-mail: mukhon@rpi.edu},
M.J. Ramsey-Musolf$^{b,e}$
\thanks{ e-mail: musolf@phys.washington.edu},
Steven J. Pollock$^{c}$
\thanks{ e-mail: pollock@lucky.colorado.edu},
J\`un L\'\i u$^{a}$
\thanks{e-mail: jliu@cs.uh.edu},
and H.-W. Hammer$^{b,d}$
\thanks{ e-mail: hammer@alph02.triumf.ca}}
\bigskip
\address{
$^a$Physics Department, Rensselaer Polytechnic Institute,
Troy, NY 12180-3590\\
$^b$Institute for Nuclear Theory, University of Washington,
Seattle, WA 98195\\
$^c$Department of Physics, University of Colorado, CB 390,
Boulder, CO 80309\\
$^d$TRIUMF, 4004 Wesbrook Mall, Vancouver, B.C.,
Canada V6T 2A3\\
$^e$Department of Physics, University of Connecticut, Storrs, CT
06260\\
}
\preprint{DOE/ER/40561-262-INT96-17-013\ and\ 
RPI-N119-98}
\title{
Parity-Violating Excitation of the $\Delta(1232)$:\\
Hadron Structure and New Physics}
\date{January 1998}
\maketitle

\begin{abstract}

We consider prospects for studying the parity-violating (PV)
electroweak excitation of the $\Delta \left( 1232\right) $ resonance
with polarized electron scattering. Given present knowledge of Standard
Model parameters, such PV experiments could allow a determination of
the $N\to\Delta$ electroweak helicity amplitudes. We discuss the
experimental feasibility and theoretical interpretability of such a
determination as well as the prospective implications for hadron
structure theory. We also analyze the extent to which a PV $N\to\Delta$
measurement could constrain various extensions of the Standard Model.

\end{abstract}

\def\sstw{{\sin^2\theta_{\scriptscriptstyle W}}}

\section{Introduction}

With the advent of continuous wave electron accelerators at Mainz and
the Jefferson Laboratory (formerly CEBAF),
studies of the hadronic electroweak response at low- and
intermediate-energy have entered a new era. One hopes to use these
facilities to develop a better understanding of the structure of
hadrons in terms of QCD quark and gluon degrees of freedom. In
addition, there exists the possibility of performing searches for
physics beyond the Standard Model (SM) of electroweak interactions.
>From both standpoints, an interesting class of observables are those
which depend on target or electron spin. Recent advances in
polarization technology have opened the way to precise measurements of
spin observables. Information gleaned from these observables provides a
more detailed probe of hadron structure and \lq\lq new physics" than
does data from spin-independent measurements alone \cite{4}.

As an example of the information which spin-observables might provide,
we consider in this paper the electroweak excitation of the $\Delta
$(1232) resonance:
\begin{equation}
N\stackrel{V}{\rightarrow }\Delta \left( 1232\right) ,
\end{equation}
where $N$ is the nucleon and $V=\gamma$, $W^\pm$, or $Z^0$.  The SM has
been tested with high precision in a variety of sectors \cite{Lang95},
so that one
knows the basics of the probe reactions extremely well. In principle,
then, one may  use these electroweak processes to study hadron
structure as it bears on the $N\to\Delta$ transition amplitudes.
Conversely, were one to have the hadronic current matrix elements
sufficiently well in hand, one might exploit this transition as a probe
of possible \lq\lq new physics" beyond the standard electroweak
theory.

The $N\to\Delta$ transition is potentially useful for either purpose,
since (i) the $\Delta \left( 1232\right) $ resonance is nicely isolated
from the plethora of other densely populated nucleon resonance states
that appear at higher energies, and (ii) it is a pure isovector,
spin-flip transition. The first of these features simplifies the
theoretical extraction of the matrix element $\langle \Delta| J_\mu |
N\rangle$ ($J_\mu$ is the appropriate electroweak current) from the
experimental observables, while the second affords one a kind of \lq\lq
filter" for selecting on various aspects of hadron structure or new
physics. For example, there has been considerable interest recently in
the strange quark content of non-strange hadrons \cite{4,squark}.
Since $s\overline{s}$ pairs contribute only to
isoscalar current matrix elements, the $N$ to $\Delta$ transition
filters out $s\overline{s}$
contributions.  Similarly, the isovector character of this transition
gives it a different sensitivity to possible contributions from
additional heavy particles not appearing in the SM than does, say, the
weak charge measured in atomic parity violation.
The $N\to\Delta$ offers the
additional advantage that it only couples strongly to one outbound
channel, viz., $N\pi $. This allows one, in principle, to treat the
unitarity issue quite rigorously\cite{1} implementing in the
theoretical analysis the constraints of the Fermi-Watson theorem
\cite{2}.

The physics of the $N\to\Delta$ transition can be probed through
a variety of processes:

\medskip
\noindent Electromagnetic (EM):
\begin{equation}
\label{eqEM}\gamma +N^a\rightarrow \Delta^a \left( 1232\right)
\rightarrow N^b+\pi^c ,
\end{equation}
\noindent Weak neutral current (NC):
\begin{equation}
\label{eqNC}\ell+N^a\rightarrow\ell+\Delta^a\rightarrow\ell+N^b+\pi^c,
\end{equation}
\noindent Weak charge changing (CC):
\begin{equation}
\label{eqCC}
\nu_l (\overline{\nu}_l)+N^a\rightarrow \ell^{-(+)}+
	\Delta ^b\rightarrow \ell^{-(+)} + N^c+\pi^d ,
\end{equation}
or the inverse CC reactions involving incident charged leptons.  Here,
$\gamma $ is a real or virtual photon, $\ell^\pm$ are charged leptons,
$\pi^i $ denotes any of the charged or neutral pion states, and $N^i$
and $\Delta^i$ are appropriate nucleon and delta states, respectively.
In addition, the axial vector transition matrix can potentially be
studied \cite{6} by the purely EM reaction
\begin{equation}
\label{2pions}e^{-}+N^a\rightarrow e^{-\prime }+\Delta^b+\pi^c
\rightarrow e^{-\prime}+ N^d+\pi^e+\pi^c,
\end{equation}
in the limit that the final state pion is soft.

The EM processes yield three resonant helicity amplitudes for the
virtual photons, $A_{1/2}^T$, $ A_{3/2}^T $ and $A_{1/2}^L$: two
transverse and one longitudinal indicated by the appropriate
superscript\footnote{The subscript $h$ indicates the absolute value of
the helicity of the virtual vector boson--nucleon system along the $\vec q$
direction (i.e. the helicity of the produced $\Delta$).}.  The
longitudinal amplitude is absent for real photons.
The primary information one has regarding the vector
helicity amplitudes has been obtained from EM processes (\ref{eqEM})
with real photons. Recent neutral pion photoproduction
experiments using polarized photons at the
Brookhaven LEGS facility\cite{Kha95} and Mainz\cite{Bec97} have determined
the vector helicity amplitudes with better than 2\% precision\cite{Dav97}.
The value of the E2/M1 ratio, extracted from the vector
helicity amplitudes, is of particular interest from the standpoint of
hadron structure theory, as it yields information on
deviations from spherical symmetry, possibly arising from tensor quark-quark
interactions. It is predicted to vary strongly with $q^2$ and reach
unity as $q^2$ becomes large\cite{Car86}. At present,
the E2/M1 ratio is poorly known beyond the photon point,
with precision decreasing as $q^2$ increases. One expects electro-production
and Compton scattering experiments at the Jefferson Lab to yield
significant improvements in precision with which one knows the E2/M1
ratio for $q^2\not=0$ as well as that of other vector current observables.
Ultimately,
the most serious obstacle to decreasing the uncertainty in the vector
amplitudes may be the theoretical problem of separating background
from resonance contributions \cite{1}.

The weak NC and CC
reactions are sensitive to the weak vector current analogues of the $A^{T,L}_h$
as well as to additional helicity amplitudes associated with the
weak axial vector current. Existing data obtained from  CC reactions
(\ref{eqCC}) provide a crude corroboration of the present knowledge
of the vector form factors \cite{18}, but do not improve
upon the precision obtained via the electromagnetic processes.
The same CC reactions also yield information on the axial vector
amplitudes and, in principle, provide a determination of the
$N\to\Delta$ axial vector form factors. However, the data
only afford a highly model-dependent determination of resonant
axial form factor parameters, and the associated uncertainties
are large \cite{18}.

There is, to date, no direct experimental neutral current data
for the process in Eq.~(\ref{eqNC}). Nevertheless, the prospects
for obtaining such data with parity-violating (PV) electron scattering
with the G0 detector at the Jefferson Lab appear to
be promising \cite{Wel97,Wel97b}. Indeed, a proposal by the G0 collaboration
to measure the PV $N\to\Delta$ transition has recently been approved 
\cite{Wel97b}. In this paper, we therefore
focus on the NC process of Eq.~(\ref{eqNC}), and
illustrate how knowledge of the NC $N\to\Delta$ helicity amplitudes
might complement existing information. In particular, we analyze
the sensitivity of the PV asymmetry to various scenarios for extending
the Standard Model and to the axial vector transition form factors
of interest to hadron structure theory. To that end, we study the
kinematic dependence of the figure of merit (FOM) for measuring
the asymmetry and identify the optimal conditions for a probe of
new physics or an extraction of the axial amplitudes. We also
estimate the contribution from non-resonant background processes
under these different conditions. Since one has no experimental means of
separating resonant from non-resonant contributions with a single PV
measurement, a theoretical subtraction of the latter is required
in any determination of resonant amplitudes. Finally, as there
exists no recent publication devoted to the NC $N\to\Delta$
transition beyond the early work of Refs. \cite{5,10}\footnote{see
also Section 4.7 of Ref. \cite{4}.} we provide
a compendium of model predictions for the various transition form
factors as well as a review of various kinematic properties and
notational conventions. Our hope is to provide
a resource for experimentalists studying the feasibility of a
PV $N\to\Delta$ measurement.

In order to guide the reader, we summarize the main conclusions of our
study here:

\begin{quotation}

(i)A determination of the axial vector response at the 25-30\%
level appears feasible with realistic run times. Moreover, lack of 
knowledge of the non-resonant background contributions does not appear to
introduce problematic theoretical uncertainties into the extraction of
$N\to\Delta$ axial vector amplitudes from the axial response.

\medskip
(ii)At leading order in electroweak couplings, the axial response
is dominated by the ratio of axial vector and vector current form factors,
$C_5^A/C_3^V$. A 25\% determination of this ratio would signficantly improve
upon information available from CC reactions, provide a new test of baryon
structure theory, and shed new light on the experimental-theoretical
discrepancies arising in the purely vector current sector. However, the extent
to which higher-order electroweak corrections cloud the extraction of
$C_5^A/C_3^V$ from the PV asymmetry remains unclear.

\medskip
(iii)As a probe of new electroweak physics, a measurement of the PV
$N\to\Delta$ transition must achieve significantly better than one percent
precision in order to compete with other low-energy semi-leptonic measurements.
At present, non-resonant background contributions are not sufficiently 
well-known to allow for a sub-one percent new physics search with this 
process.

\end{quotation}

Our discussion of these points in the remainder of the paper is organized as
follows.In Section II, we discuss some of the general features of the PV 
asymmetry and the way in which various types of physics enter. Section III 
gives a review of kinematic properties. In Section IV, we outline the
formalism for analyzing the appropriate response functions and
show how various pieces of the asymmetry depend on different aspects
of these response functions. In Section V, we discuss the sensitivity
of the asymmetry to different types of \lq\lq new physics", and in
Section VI we review the physics issues associated with the axial
response. Section VII treats the theoretical uncertainties arising
in the interpretation of the asymmetry, including those associated
with backgrounds. In Section VIII, we discuss experimental considerations,
including the FOM and prospective statistical uncertainty as a function
of kinematics. Section IX summarizes our results.

\section{PV Asymmetry: General Features}

The NC helicity amplitudes can be obtained
from the parity-violating (PV) helicity-difference, or
\lq\lq left-right", asymmetry for the scattering of longitudinally
polarized electrons from a nucleon target \cite{4,5,10}:
\begin{equation}
\label{alr1}
A_{LR}={{N_+ - N_-}\over{N_++N_-}}=
\frac{-G_\mu}{\sqrt{2}}\frac{|q^2|}{4\pi \alpha }\left[
\Delta^\pi_{(1)} + \Delta^\pi_{(2)}+\Delta^\pi_{(3)}\right] ,
\end{equation}
where $N_{+}$ ($N_{-}$) is the number of detected, scattered electrons
for a beam of positive (negative) helicity electrons; $q^2$ is the
square of the four-momentum transfer to the target; $\alpha $ and
$G_\mu$ are, respectively, the electromagnetic fine structure constant
and the Fermi constant measured in $\mu$-decay.

The quantities $\Delta^\pi_{(i)}$ ($i=1,\ldots , 3$) denote the three
primary contributions to the asymmetry. Specifically, one has
\begin{equation}
\label{delpi1}\Delta^\pi_{(1)}=g_{A}^e{\xi_V^{T=1}} \ \ \ ,
\end{equation}
which includes the entire resonant hadronic vector current contribution to the
asymmetry. Here, $g_{A}^e$ is the axial vector electron coupling to the $Z^0$
and ${\xi_V^{T=1}}$ is the isovector hadron-$Z^0$ vector current
coupling \cite{4,mus92a}. At tree level in the SM, these lepton and hadron
couplings take on the values 1 and
$2(1-2{\sin^2\theta_{\scriptscriptstyle W}})$, respectively. At leading
order, this term contains no dependence on hadronic form factors, owing
to a cancelation between terms in the helicity-dependent and
helicity-independent cross sections. The quantity $\Delta^\pi_{(2)}$ contains
residual contributions from non-resonant, hadronic vector current background
transitions, not taken into account via $\Delta^\pi_{(1)}$. The third term,
$\Delta_{(3)}^\pi$, involves the axial-vector $N\to\Delta$ coupling:
\begin{equation}
\label{delpi3}\Delta^\pi_{(3)}\approx g_{V}^e{\xi_A^{T=1}}
F(q^2,s)\ \ \ ,
\end{equation}
as well as hadronic axial vector background contributions.
At tree level in the Standard Model, these couplings take on the values
$g_V^e=-1+4{\sin^2\theta_{\scriptscriptstyle W}}$
and $\xi_A^{T=1}=-2$. The function
$F(q^2, s)$ involves a ratio of PV and parity-conserving (PC)
electroweak response functions (see Section IV).
The variable $s$ is the square of
the total energy in the center of mass frame.  In writing down the RHS
of Eq.~(\ref{delpi3}), we have ignored non-resonant axial
vector contributions for simplicity.

The physics of interest lies in $\Delta^\pi_{(1)}$ and
$\Delta^\pi_{(3)}$.  The background term $\Delta^\pi_{(2)}$ introduces
a theoretical uncertainty into the extraction of the other two terms
from $A_{LR}$.  A precise measurement of the first term,
$\Delta^\pi_{(1)}$, would provide a window on physics beyond the SM.
This term is, at tree level, independent of any hadronic matrix
elements and involves only the product of NC electron and isovector NC
quark couplings. The latter, ${\xi_V^{T=1}}$, has never been determined
independently from the other hadronic vector neutral current
couplings.  In the discussion below, we consider the sensitivity of
${\xi_V^{T=1}}$ to two types of new physics: \lq\lq oblique
corrections", which involve corrections to the $Z^0$ and $W^\pm$
propagators from new heavy particles, and \lq\lq direct" contributions,
such as those from tree-level exchanges of additional neutral gauge
bosons or leptoquarks, which manifest themselves as new four-fermion
contact interactions. The $Z^0$-pole observables are relatively
insensitive to direct interactions, in contrast to the situation with
low-energy observables like $A_{LR}(N\to\Delta)$.

As a benchmark, we consider the constraints on new physics which a one
percent determination of ${\xi_V^{T=1}}$ might yield and compare these
prospective constraints with those obtainable from other low-energy NC
observables. We also discuss the experimental conditions under which a
one percent determination might be made, along with the theoretical
uncertainties which may enter at that level. We find that a one percent
knowledge of ${\xi_V^{T=1}}$ would provide constraints roughly
comparable to those presently obtained from atomic PV.  A fairly
demanding experimental setup, with 1000 hours of moderately high energy
($\gtrsim$ 3 GeV) CEBAF\footnote{The accelerator and three experimental
halls at the Jefferson Lab are collectively titled the Continuous Electron
Beam Accelerator Facility (CEBAF).}
beam running at forward angles might be able to
achieve this level, if the non-resonant backgrounds can be understood
at roughly $\pm$25\% levels or better, and the axial contributions can
be understood at roughly $\pm$30\% levels or better. For more
quantitative details, see section VII.

The third term, $\Delta^\pi_{(3)}$, is interesting from the standpoint
of hadron structure. To a good approximation, the function $F(q^2, s)$
contained in $\Delta^\pi_{(3)}$ is proportional to the
ratio of two transition form factors: $C^A_5/C^V_3$, where $V$ ($A$)
correspond to the hadronic vector (axial vector) current. This ratio is
the off-diagonal analog of the $G_A/G_V$ ratio extracted from neutron
$\beta$-decay. A measurement of $\Delta^\pi_{(3)}$ could
correspondingly provide an opportunity to test low-energy consequences
of chiral symmetry, such as the off-diagonal Goldberger-Treiman
relation and its (small) chiral corrections \cite{HHM}.

A separate determination of $C^A_5$ is also of interest in light of
experimental results\cite{Kha95,Bec97}
for the vector current transition form
factors from photoproduction of pions. In the
latter case, for example, one finds a significant difference with
quark model predictions\cite{Dav97} and lattice calculations\cite{Lei92}.
For the magnetic transition form factor, the
disagreement is at the 30\% level, while for the transition charge
radius, the disagreement arises at the 20\% level \cite{1}. A variety
of scenarios have been proposed to account for this discrepancy, such
as the role of the meson clouds around the quark core or a change
of the value of the quark magneton\cite{i}. A new determination of
$C^A_5$ could provide an additional test of lattice and quark
model calculations and of the recipes proposed for correcting
the vector form factor discrepancies. As we discuss below, a
30\% determination of $C_5^A$ from a measurement of $A_{LR}(N\to\Delta)$
appears to be within the realm of feasibility.

Whether these benchmarks for ${\xi_V^{T=1}}$ and
$C^A_5/C^V_3$ can be realized depends, in part, on the degree to which
theoretical uncertainties entering the interpretation of
$A_{LR}(N\to\Delta)$ are sufficiently small. The most serious
uncertainties appear in two guises: (i) background contributions,
contained in $\Delta^\pi_{(2)}$, and (ii) hadronic contributions to
electroweak radiative corrections, which enter both $\Delta^\pi_{(1)}$
and $\Delta^\pi_{(3)}$. A recent analysis of the background
contributions was reported in Ref.~\cite{11}. We recast that analysis
into the formalism of the present study and extend the estimates
of Ref.~\cite{11} to cover the full range of Jefferson Lab kinematics.
It appears that, under appropriate kinematic conditions,
our estimate of this background
contribution is sufficiently small -- even given fairly large uncertainties
in its absolute value -- to allow a reasonable determination of
$C^A_5$/$C^V_3$ ratio from PV asymmetry measurements. Significant
improvements in knowledge of $\Delta^\pi_{(2)}$, however, would be
needed to permit a one percent determination of $\xi_V^{T=1}$.

The situation regarding radiative correction uncertainties is less
clear. Two classes of hadronic effects in these corrections require
further study: two-boson exchange \lq\lq dispersion corrections"
\cite{mus93a} and
corrections induced by PV quark-quark interactions in the hadronic
vertex. The former enter the analysis of all three of the
$\Delta^\pi_{(i)}$ while the latter contribute to $\Delta^\pi_{(3)}$ only.
We especially highlight the hadronic PV effects since,
in the case of elastic PV
electron scattering from the proton, they introduce considerable
theoretical uncertainties into the axial vector part of the asymmetry
\cite{mus90}. Although an estimate of these hadronic PV corrections
goes beyond the scope of the present paper, we emphasize the importance
of performing such an estimate when seeking to extract $C^A_5/C^V_3$
from $\Delta^\pi_{(3)}$.

\section{Kinematics}

In this section, we review the basic kinematics of the process in
Eq.~(\ref{eqNC}) -- depicted in Fig. 1 --  which we rewrite as
\begin{equation}
\label{eq16}e^{-}\left( k\right) +N\left( p\right) \rightarrow
e^{-\prime }(k^{\prime}) +\Delta \left( p_\Delta \right) \rightarrow
e^{-\prime }\left( k^{\prime }\right) +
        N^{\prime }\left( p^{\prime }\right) +\pi \left( p_\pi \right) ,
\end{equation}
omitting the superscripts. The kinematic variables have been explicitly
indicated by the four-momenta of the particles in brackets.
Expressing quantities in the laboratory frame,
\begin{equation}
\label{eq17}s=\left( k+p\right) ^2,\quad q=p_\Delta -p=k-k^{\prime },\quad
p_\Delta =p^{\prime }+p_\pi ,
\end{equation}
where ${\bf p}=0$, and
\begin{equation}
\label{eq18}s=k^2+2k\cdot p+p^2=m^2+2M\epsilon+M^2,
\end{equation}
$\epsilon$ being the incoming electron energy, $m$ and $M$ are the electron
and nucleon
masses. We assume the electrons are ultra-relativistic, and
henceforth ignore $m$. In this case, one has
\begin{equation}
\label{eq19}\epsilon \ =\  \frac{s-M^2}{2M}.
\end{equation}
The outgoing electron lab energy $\epsilon^{\prime }$ is given by
\begin{equation}
\label{eq20}\epsilon^{\prime }=\epsilon+M-\sqrt{M_\Delta ^2+{\bf p}_\Delta ^2
}=\epsilon+M-\sqrt{
M_\Delta ^2+\left( {\bf k}-{\bf k}^{\prime }\right) ^2},
\end{equation}
where
\begin{equation}
\label{eq21}\left( {\bf k}-{\bf k}^{\prime }\right) ^2=
\epsilon^2+\epsilon^{\prime 2}-2\epsilon\epsilon^{\prime }\cos\theta ,
\end{equation}
$\theta $ being the lab scattering angle. Solving for $\epsilon^{\prime }$
gives
\begin{equation}
\label{eq22}\epsilon^{\prime }=\frac{2M\epsilon+M^2-M_\Delta ^2}{2M+2\epsilon
\left( 1-\cos\theta
\right) }.
\end{equation}
Writing the four-momentum transfer squared as $Q^2=-q^2$
(a positive quantity in this process),
\begin{eqnarray}
Q^2\equiv -q^2 \ &=&\ 2\epsilon\epsilon^{\prime }\left( 1-\cos\theta \right)
=
4\epsilon\epsilon^{\prime }\sin^2\theta/2.\label{eq24} \\
&=&\ \frac{\left( s-M^2\right) \left( s-M_\Delta ^2\right) }
{ M^2+\left( s-M^2\right) \sin^2\theta/2} \sin^2\theta/2.
\label{eq25}
\end{eqnarray}
Since $Q^2\geq 0$, we have lower bounds for $s$ and
$\epsilon$ for the process of Eq.~(\ref{eq16}):
\begin{equation}
\label{eq26a}s\geq M_\Delta ^2,
\end{equation}
\begin{equation}
\label{eq26b}\epsilon\geq \frac{M_\Delta ^2-M^2}{2M}.
\end{equation}
On the other hand, from Eq.~(\ref{eq25}), we can
rewrite $\sin^2\frac \theta 2$ as
\begin{equation}
\label{eq27}\sin^2\theta/2=\frac{M^2Q^2}{\left( s-M^2\right) \left(
s-M_\Delta ^2-Q^2\right) }.
\end{equation}
Since $0\leq\sin^2\theta/2\leq 1$, Eq.~(\ref{eq27}) leads to a
bound for $Q^2$ for our reaction of interest
\begin{equation}
\label{eq28}Q^2\leq \frac{\left( s-M^2\right)
\left( s-M_\Delta ^2\right) }s.
\end{equation}

Other useful kinematics identities are summarized as follows:
the energy available in the nucleon-gauge boson ($\gamma$ or $Z^0$)
center of mass (CM) frame is
$W\equiv\sqrt{p_\Delta^2}$. (The above equations assumed a narrow $\Delta$,
henceforth we use W instead of $M_\Delta$.)
The energy of the gauge boson
in the CM frame is
\begin{equation}
\label{kin1}
q^0\ =\ {{W^2-Q^2-M^2}\over{2W}}.
\end{equation}
The gauge boson three momentum is given by
\begin{equation}
\label{kin2}
q^*\ \equiv \    \left|\,{\vec q}\,(CM)\,\right| \ =\
\frac{M}{W}\left|\,{\vec q}\,(lab)\,\right|\ =\
\sqrt{{\left(Q^2+(M+W)^2\right)\left(Q^2+(M-W)^2\right)}\over{4 W^2}},
\end{equation}
and another useful identity is
\begin{equation}
\label{kin3}
W+M-q^0\ =\ M+P_0\ =\ {{Q^2+(M+W)^2}\over{2 W}},
\end{equation}
where $P_0$ is the proton energy in the CM frame.

\section{$N$ to $\Delta$ response functions}

To calculate the differential cross-section and PV asymmetry for the
process $N( \vec{e},e^{\prime }) \pi N^{\prime }$, we consider the
exchange of one neutral gauge boson between the lepton and hadron
vertices. The general expression for the differential cross-section can
be written directly in terms of electroweak lepton and hadron response
tensors.  For details of the Lorentz structures of these tensors, we
refer the reader to the older papers of Adler \cite{9} and more recent
discussions in e.g. Refs.~\cite{4} and ~\cite{11}.  Considering first the
electromagnetic terms only, and integrating over the outgoing hadronic
momenta, the double differential cross-section can be written as
\cite{4,12}
\begin{equation}
\label{sigtot}
\frac{d^2\sigma }{d\Omega d\epsilon^{\prime }}=
\sigma_M W^{EM}(|\vec q\,|, \omega, \theta)\ \ \ ,
\end{equation}
where $\omega$ and $|\vec q\,|$ are the energy and three-momentum
transferred to the hadronic system in the laboratory frame, $\sigma_M$ is
the Mott cross section or scattering from a point-like target
\begin{equation}
\sigma_M=\left({\alpha\cos\theta/2\over 2\epsilon\sin^2\theta/2}\right)^2
\ \ \ ,
\end{equation}
and $W^{EM}$ is the EM transition response function
\begin{equation}
\label{wem} W^{EM}(|\vec q\,|,\omega,\theta)=v_L R^L(|\vec q\,|,\omega)+
	v_T R^T(|\vec q\,|, \omega)\ \ \ .
\end{equation}
In Eq.~(\ref{wem}) the two kinematic factors are
\begin{eqnarray}
v_L&=& \left(Q^2/|\vec q\,|^2\right)^2 \\
v_T&=&{1\over 2}\left(Q^2/|\vec q\,|^2\right)+\tan^2{\theta/2}
\end{eqnarray}
The longitudinal ($L$) and transverse ($T$) response functions are
given, using the conventions of Zucker \cite{13} by
\begin{eqnarray}
\label{zi}
R^L(|\vec q\,|, \omega)&=&
\left({W^2\over 2 |\vec q\,| M^2}{W^2\over M^2}\right)
\sum_{J_\pi }\left( 2J+1\right)
  \left| T_C^{J_\pi}\right|^2 \\
R^T(|\vec q\,|, \omega)&=&
\left({W^2\over 2 |\vec q\,| M^2}\right)
\sum_{J_\pi }
\left( 2J+1\right)
\left( \left| T_{1/2}^{J_\pi}\right| ^2+
          \left| T_{3/2}^{J_\pi}\right| ^2\right)  \label{rt}
\end{eqnarray}
where $J(\pi)$ indicates the spin (parity) channel for the $\pi N$
system, $T_C^{J_\pi}$ is a Coulomb amplitude (corresponding to helicity
$h=1/2$), and the $T_h^{J_\pi}$, $h=1/2, 3/2$ are transverse helicity
amplitudes.

The PV asymmetry involves a ratio of two response functions:
\begin{equation}
\label{alr2}
A_{LR}(N\to\Delta)={G_\mu Q^2\over 2\sqrt{2}\pi\alpha}{W^{PV}\over W^{EM}}
\end{equation}
where $W^{PV}$ is a PV response function arising from the interference
between the EM and PV NC amplitudes. Its structure is similar to that of
$W^{EM}$:
\begin{equation}
W^{PV}(|\vec q\,|, \omega, \theta)=v_L R^L_{AV}(|\vec q\,|, \omega)+
v_T R^T_{AV}(|\vec q\,|, \omega)+v_{T'} R^{T'}_{VA}(|\vec q\,|, \omega)
\end{equation}
where the response functions $R^{L,T}_{AV}$ are the EM-NC interference
analogues of the EM response functions and where the third term arises from
an interference between the EM vector current and weak neutral axial
vector current\cite{TWD88,mus92a}. The corresponding kinematic coefficient is
\begin{equation}
v_{T'}=\tan\theta/2\ \sqrt{Q^2/|\vec q\,|^2+\tan^2{\theta/2}}
\end{equation}
and the interference response functions are
\begin{mathletters}  
\begin{eqnarray}
R^L_{AV}(|\vec q\,|, \omega)&=&
-{1\over2}g_A^e
\left({W^2\over 2 |\vec q\,| M^2}{W^2\over M^2}\right)
\sum_{J_\pi }
\left( 2J+1\right)\ \hbox{Re}\ {\tilde T_C}^{J_\pi } T_C^{J_\pi \ast} \\
R^T_{AV}(|\vec q\,|, \omega)&=&
-{1\over2}g_A^e
\left({W^2\over 2 |\vec q\,| M^2}\right)
\sum_{J_\pi }
\left( 2J+1\right)\ \hbox{Re}\left({\tilde T}_{1/2}^{J_\pi}
T_{1/2}^{J_\pi\ \ast}+
         {\tilde T}_{3/2}^{J_\pi } T_{3/2}^{J_\pi\ \ast}\right) \\
R^{T'}_{VA}(|\vec q\,|, \omega)&=&
-g_V^e
\left({W^2\over 2 |\vec q\,| M^2}\right)
\sum_{J_\pi }
\left( 2J+1\right)\ \hbox{Re}\left(
{\tilde U}_{1/2}^{J_\pi} T_{1/2}^{J_\pi\ \ast} -
{\tilde U}_{3/2}^{J_\pi} T_{3/2}^{J_\pi\ \ast}\right)
\label{zx}
\end{eqnarray}
\end{mathletters}
where the multipoles ${\tilde T}$ (${\tilde U}$) now involve
projections of the neutral vector (axial vector) current\cite{13}.  The
subscripts $AV$ ($VA$) indicate that the corresponding PV NC amplitude
arises from an axial vector (vector) coupling of the $Z^0$ to the
electron and a vector (axial vector) coupling at the hadronic vertex.

The physics of the asymmetry is governed by the \lq\lq hadronic ratio"
$W^{PV}/W^{EM}$ appearing in Eq.~(\ref{alr2}).  The vector NC
multipoles arising in the numerator of this ratio
can be related in a straightforward manner to those
involving the isovector ($T=1$) and isoscalar ($T=0$) components of the
EM current.  The axial vector multipoles can decomposed into their
SU(3) components. To that end, we follow the notation of
Refs. \cite{4,mus92a} and
write
\begin{mathletters}  
\begin{eqnarray}
\label{emnc}
J^{EM}_\mu&=&J^{EM}_\mu(T=1)+ J^{EM}_\mu(T=0) \\
\label{emncb}
J^{NC}_\mu&=&\xi_{V}^{T=1} J^{EM}_\mu(T=1)+\sqrt{3} \xi_{V}^{T=0}
J^{EM}_\mu(T=0)+\xi_{V}^{(0)} V^{(s)}_\mu\\
\label{emncc}
J^{NC}_{\mu 5}&=&\xi_A^{T=1} A_\mu^{(3)}+\xi_A^{T=0} A_\mu^{(8)}
+\xi_A^{(0)} A_\mu^{(s)}
\end{eqnarray}
\end{mathletters}
Here, the $\xi_{V}^{(a)}$ are electroweak hadronic couplings,
$A_\mu^{(a)}$ are components of the octet of axial vector
currents, and $V^{(s)}_\mu$ ($A^{(s)}_\mu$) is the vector (axial vector)
current associated with strange quarks only. In arriving
at the foregoing expressions, we have omitted the $c$-, $b$- and
$t$-quark contributions. At tree level in the Standard Model, the
electroweak couplings are given by
\begin{eqnarray}
\label{tree}
\xi_{V}^{T=1}&=&2(1-2\sin^2\theta_W) \\
\sqrt{3}\xi_{V}^{T=0}&=&-4\sin^2\theta_W  \nonumber \\
\xi_{V}^{(0)}&=&-1 \nonumber\\
\xi_{A}^{T=1}&=&-2\nonumber \\
\sqrt{3}\xi_{A}^{T=0}&=& 0 \nonumber \\
\xi_{V}^{(0)}&=&1 \nonumber
\end{eqnarray}

It is useful to exploit Eqs.~(\ref{emnc},\ref{emncb})  and relate the NC vector
multipoles to the corresponding EM multipoles. To that end, we
decompose the various amplitudes into their isospin components.  In the
EM case, the $p\to N\pi$ amplitudes are as follows, where we have
suppressed all indices referring to spin, parity, and Lorentz structure
for simplicity, retaining {\it only} the isospin structure:
\cite{4,12,walker}:
\begin{eqnarray}
\label{36}
T_{n\pi^{+}}\equiv T_h^{J_\pi}(n\pi^+)&=&
- \sqrt{2}T^{IS}+T^{T=1/2}+{1\over\sqrt2}T^{T=3/2} , \\
T_{p\pi ^0}\equiv T_h^{J_\pi}(p\pi^0)&=&
T^{IS}-{1\over\sqrt2}T^{T=1/2}+T^{T=3/2}, \nonumber
\end{eqnarray}
The amplitude $T^{IS}$ is isoscalar, while $T^{T=1/2}$ and
$T^{T=3/2}$ are
linearly independent isovector amplitudes going to isospin $1/2$ and
$3/2$ respectively.  For pure $\Delta$ production, of course,
$T^{IS}=T^{T=1/2}=0$.  We can similarly decompose the weak vector or axial
vector amplitudes.  The vector NC amplitudes are then straightforwardly
related to the EM and strange quark amplitudes using
Eqs.~(\ref{emnc},\ref{emncb}):
\begin{eqnarray}
\label{38}
{\tilde T}^{IS}&=&\sqrt{3}\xi_{V}^{T=0} T^{IS}+\xi_{V}^{(0)} T^{(s)},\\
{\tilde T}^{T=1/2\  (3/2)}&=&\xi_{V}^{T=1} T^{T=1/2 \ (3/2)}. \nonumber
\end{eqnarray}
where $T^{(s)}$ is an amplitude from the
strange-quark vector current.

Using these relations, we examine the weak vector and electromagnetic
interference terms which generate $W^{PV}$:
\begin{eqnarray}
\label{39}
\left( \tilde T^{*}T\right) _{p\pi ^0}+
\left( \tilde T^{*}T\right) _{n\pi ^{+}} \ &=&\
\xi_V^{T=1}\left(|T_{p\pi^0}|^2+|T_{n\pi^+}|^2\right) \\
&& \ \  + \ \left[(\sqrt3\xi_V^{T=0}-\xi_V^{T=1})
	T^{IS*}+\xi_V^{0}T^{(s)*}\right]
\left(T_{p\pi^0}-\sqrt2T_{n\pi^+}\right). \nonumber
\end{eqnarray}
This expression is derived using Eqs.~(\ref{36}) and (\ref{38}), adding and
subtracting $\xi_V^{T=1}T^{IS}$ to the isovector terms in
order to pull out the overall isovector factor of $\xi_V^{T=1}$ in the
first term.  The first term is then exactly proportional to the
unpolarized EM cross section. The second term gives corrections arising
entirely from non-resonant backgrounds, as the quantity $T_{p\pi^0}
-\sqrt{2}T_{n\pi^+}$  contains no $T=3/2$ component.
We find these corrections to
be small in the region of the $\Delta$ resonance.
Using Eq.~(\ref{39}), we can
rewrite the asymmetry for inclusive pion production from the proton
target as given in Eq.~(\ref{alr1}). The origin of the
various terms $\Delta^\pi_{(a)}$ is now transparent. The first term,
$\Delta^\pi_{(1)}$ arises from the part of the EM-NC interference which
is proportional to the total EM response (the first term on the RHS of
Eq.~(\ref{39}) above).  Indeed, the EM response is proportional to
$|T_{p\pi^0}|^2+|T_{n\pi^+}|^2$, summed over the appropriate spins
and parities. The corresponding hadronic matrix elements thus
cancel out entirely from this part of the asymmetry, leaving only the
dependence on $\xi_{V}^{T=1}$, as indicated in Eq.~(\ref{delpi1}).

The contribution $\Delta^\pi_{(2)}$ arises from those non-resonant
vector current background amplitudes which do not cancel from
the hadronic ratio. Neglecting contributions from $T^{(s)}$,\footnote{Although
we expect these contributions to be small, estimates of their prospective
magnitude remain to be computed. Lack of knowledge in the strange quark
contributions constitutes one source of theoretical uncertainty in the
backgrounds.}
one has
\begin{eqnarray}
\label{43}
W^{EM}\Delta^\pi_{(2)}\ =\
+g_A^e&&(\sqrt3\xi_V^{T=0}-\xi_V^{T=1})
{W^2\over2M^2 |\vec q\,|} \sum_{J_\pi}\ {\hbox{Re}}\ (2J+1)\times \\
&& \ \biggl[
v_L{W^2\over M^2}
T_{C}^{IS*} \left(3T_{C}^{IS}     -(3/\sqrt2)T_{C}^{T=1/2} \right)\nonumber \\
&& \ + v_T \biggl\{
T_{1/2}^{IS*} \left(3T_{1/2}^{IS} -(3/\sqrt2)T_{1/2}^{T=1/2}
\right)\nonumber\\
&& \ \qquad\
 + T_{3/2}^{IS*} \left( 3T_{3/2}^{IS} -(3/\sqrt2)T_{3/2}^{T=1/2} \right)
			\biggr\} \biggr] \nonumber
\end{eqnarray}
where we have continued to suppress the spin-parity indices and
where the linear combination of electroweak couplings
$\sqrt3\xi_V^{T=0}-\xi_V^{T=1}$ is just twice the $Z^0$-neutron
coupling which takes on the value $-1$ at tree level\footnote{
An equivalent expression, written in terms of transverse electric and
magnetic (and longitudinal) multipoles, rather than helicity
multipoles, can be found in section 4.7 of ref.~\cite{4}.}.
In general one must investigate the corrections introduced by these terms.
We postpone a detailed discussion of their importance to Section VII.

Finally, the axial vector hadronic NC contributes via $\Delta^\pi_{(3)}$:
\begin{eqnarray}
\label{45}
W^{EM}\Delta^\pi_{(3)}\  = \
2 g_V^e {W^2\over2M^2 |\vec q\,|}
\ \sum_{J_\pi}Re\ (2J+1) v_{T'} &&
\biggl\{\left( {\tilde U}_{1/2}T_{1/2}^*-
		{\tilde U}_{3/2}T_{3/2}^*\right)_{p\pi^0}\\
&&\ \  +\left( {\tilde U}_{1/2}T_{1/2}^*-
		{\tilde U}_{3/2}T_{3/2}^*\right)_{n\pi^+}\biggr\} \nonumber
\end{eqnarray}
The isospin structure of the axial-vector term $\Delta_{(3)}^\pi$ is
not explicitly decomposed in Eq.~(\ref{45}), as there is no EM analog
from which to extract information.

It is worth noting that
the value of $g_V^e= \left(-1+4\sin^2\theta_W\right)
\approx -0.1$ suppresses the contribution from $\Delta_{(3)}^\pi$
relative to that from $\Delta_{(1,2)}^\pi$, for which the product of
leptonic and hadronic NC couplings is of order unity.
Thus, a measurement of the axial vector term is intrinsically rather difficult.
Moreover, in contrast to the
situation with elastic PV electron scattering, the hadronic axial
vector contribution to $A_{LR}(N \to\Delta)$ does not vanish at forward
angles. It is straightforward to show, using the relations of Section
II, that for $\theta\to 0$ and $Q^2\to 0$ one has $v_L/v_T\to 0$, but
\begin{equation}
{v_{T'}\over v_T}\rightarrow {\epsilon^2-\epsilon^{\prime\ 2}
	\over \epsilon^2+\epsilon^{\prime\ 2}} \not= 0.
\end{equation}
In the limit of large incoming energy, however, the ratio
${v_{T'}/ v_T}$ vanishes like
$1/\epsilon$, and the axial contribution to the asymmetry,
$\Delta_{(3)}^\pi$ becomes insignificant compared to $\Delta_{(1)}^\pi$,
which stays fixed.  For purposes of determining $\Delta^\pi_{(1)}$, then,
going to forward angles and higher energies allows one to minimize
the impact of uncertainties associated with the axial response.

>From the standpoint of hadron structure studies, the axial contribution
$\Delta^\pi_{(3)}$ is of particular interest. It is useful to express
this quantity in terms of the Adler form factors \cite{5,9,14},
\begin{eqnarray}
\label{46}
<\Delta \left( p^{\prime }\right)&& \left| V_\lambda ^3\right| N\left(
p\right) >\ =\
iN_0 \overline{U}_\nu \left( p^{\prime }\right) \times \nonumber\\
&&
\Biggl[\delta _{\nu \lambda }\left(\frac{M_\Delta +M}MC_3^V-
\frac{p^{\prime }\cdot q}{M^2}C_4^V-\frac{pq}{M^2} C_5^V+C_6^V\right)
   \nonumber\\
&&\quad   +ip_\nu \gamma_\lambda
\left(-\frac{C_3^V}M\right)
 +p_\nu \left( p+p^{\prime }\right)_\lambda
\left(-\frac{C_4^V+C_5^V}{2M^2}
\right) -p_\nu q_\lambda \frac{C_4^V-C_5^V}{2M^2}\Biggr]
\gamma_5u \left( p\right),
\end{eqnarray}
\begin{eqnarray}
\label{47}
<\Delta \left( p^{\prime }\right)&& \left| A_\lambda ^3\right| N\left(
p\right) >=iN_0
\overline{U}_\nu \left( p^{\prime }\right)\times \nonumber\\
&&
\Biggl[\delta _{\nu \lambda
}\left( -
\frac{M_\Delta -M}MC_3^A
+ \frac{p^{\prime }\cdot q}{M^2}C_4^A-C_5^A\right)
    \nonumber \\
&&\quad  +ip_\nu \gamma _\lambda \frac{C_3^A}M+p_\nu
\left( p+p^{\prime }\right)_\lambda \frac{C_4^A}{2M^2}-
p_\nu q_\lambda \left(-\frac{C_4^A+2C_6^A}{2M^2} \right)
 \Biggr]u\left( p\right) ,
\end{eqnarray}
where the baryon spinors are defined in the usual way.  Connecting to
the amplitudes of Eqs.~(\ref{zi}-\ref{zx}), considering only the
resonant J=3/2, parity $+$ terms in the vicinity of the $\Delta(1232)$,
\begin{equation}
\label{48}
T_{3/2}^{J_\pi=3/2+,\ T=3/2}=f\left( W\right) N_{RS}\ q^{*}\left[
 \left( \frac{W+M}M \right) C_3^V+\frac{Wq^0}{M^2}
 \left( C_4^V+C_5^V\right)+\frac{Q^2}{M^2}C_5^V\right] ,
\end{equation}
\begin{eqnarray}
\label{49}T_{1/2}^{J_\pi=3/2+,\ T=3/2}&=& \nonumber \\
\sqrt{\frac 13}f
\left( W\right)& & N_{RS}\ q^{*}\left[
 -\left( \frac{P_0+M-q^0}M\right) C_3^V+\frac{Wq^0}{M^2}
 \left( C_4^V+C_5^V\right)+\frac{Q^2}{M^2}C_5^V\right] ,
\end{eqnarray}
\begin{equation}
\label{50}U_{3/2}^{J_\pi=3/2+,\ T=3/2}=f\left( W\right) N_{RS}
\left( P_0+M\right)
\left[ \left(
\frac{W-M}M\right) C_3^A+\frac{Wq^0}{M^2}C_4^A+C_5^A\right] ,
\end{equation}
\begin{equation}
\label{51}U_{1/2}^{J_\pi=3/2+,\ T=3/2}=\sqrt{\frac 13}f\left( W\right) N_{RS}
\left( P_0+M\right)
\left[ \left( \frac{q^0-P_0+M}M\right) C_3^A+\frac{Wq^0}{M^2}
C_4^A+C_5^A\right] ,
\end{equation}
following the notations of Zucker \cite{13}, and Schreiner and von
Hippel \cite{14}.  The function
$f(W)$ is a Breit-Wigner line shape,
whose square is normalized
to unit area in the narrow width limit \cite{14},  and the
normalization factor $N_{RS}$ is
\begin{equation}
\label{49'}
N_{RS}\equiv -i\sqrt{q^*\over{ 4W(P_0+M)}}.
\end{equation}

In order to highlight the physics governing the behavior of
$\Delta^\pi_{(3)}$, it is useful to derive an approximate expression
for the function $F(q^2, s)$. To that end, we first drop the background
contributions to $\Delta^\pi_{(3)}$ and
assume that
the vector current component of the $N\to\Delta$ transition is dominated
by the magnetic dipole amplitude.
Under the latter assumption, one has \cite{paper1}
\begin{equation}
\label{57}
C_5^V\left( q^2\right) \approx 0,\,C_4^V\left( q^2\right)
\approx -\frac M{M_\Delta }C_3^V\left( q^2\right) .
\end{equation}
Omitting all kinematical factors common to the numerator and the
denominator in the
asymmetry equation~(\ref{alr2})
we can rewrite the $T_h^{J_\pi=3/2+,\ T=3/2}$ as
\begin{equation}
\label{58}T_{3/2}^{J_\pi=3/2+,\ T=3/2}\sim q^{*}
\left[ \frac{W+M}MC_3^V-\frac MW\frac{Wq^0}{M^2}
C_3^V\right] =q^{*}\frac{W+M-q^0}MC_3^V,
\end{equation}
\begin{equation}
\label{58'}T_{1/2}^{J_\pi=3/2+,\ T=3/2}\sim \frac 1{\sqrt{3}}q^*
\left[ -\frac{P_0+M-q^0}MC_3^V-\frac MW\frac{Wq^0}{M^2}C_3^V\right]
 =\frac{-1}{\sqrt{3}}q^{*}\frac{P_0+M}MC_3^V,
\end{equation}
with $W=M_\Delta $ at the peak of the $\Delta \left( 1232\right) $,
resulting in a total cross section proportional to
\begin{eqnarray}
\label{59}
\left| T_{1/2}^{J_\pi=3/2+,\ T=3/2}\right|^2+
\left|T_{3/2}^{J_\pi=3/2+,\ T=3/2}\right|^2&\sim& \nonumber \\
\frac{q^{*2}}{M^2} \left(C_3^V\right)^2\bigl[\frac 13\left(P_0+M\right)^2
  &+&\left(W+M-q^0\right)^2\bigr]
=\frac{{q^*}^2}{M^2} \frac{4(P_0+M)^2}{3} {C_3^V}^2 .
\end{eqnarray}

Considering now the axial component of $W^{PV}$, we note that
$C_3^A$ vanishes in the SU(6) limit and that it takes on small
values in nearly all model calculations (see Table I). Setting
$C_3^A=0$ in the ${\tilde U}_h^{J_\pi=3/2+,\ T=3/2}$ yields
\begin{eqnarray}
\label{54}{\tilde U}^{J_\pi=3/2+,\ T=3/2}_{1/2}T^{J_\pi=3/2+,\ T=3/2}_{1/2}
&-&{\tilde U}^{J_\pi=3/2+,\ T=3/2}_{3/2}T^{J_\pi=3/2+,\ T=3/2}_{3/2}
\ \sim \nonumber \\
& & \nonumber \\
\ C_3^V \left( \frac{Wq^0}{M^2}C_4^A+C_5^A\right)
q^*(P_0+M)& &
\left(\frac{-(P_0+M)}{3M}-\frac{(W+M-q^0)}{M}\right) \nonumber\\
=\ \ C_3^V C_5^A\left(1+\frac{W q^0}{M^2}\frac{C_4^A}{C_5^A}\right)& &
\left(\frac{-4}{3}\frac{q^*}{M}\right)(P_0+M)^2.
\end{eqnarray}
Taking the ratio of Eqs.~(\ref{54}) to (\ref{59}) and including the
appropriate kinematic factors gives for the axial contribution
to the asymmetry
\begin{equation}
\label{63}\Delta^\pi_{(3)} =g_{V}^e\xi_A^{T=1}\frac{C_5^A}{C_3^V}\left[
1+\frac{Wq^0}{M^2}\frac{C_4^A}{
C_5^A}\right] {\cal P},
\end{equation}
where ${\cal P}$ is the kinematic function
\begin{equation}
\label{64}
{\cal P}=\frac{v_{T'}}{v_T}\frac{W}{|\vec q\,|}\ \ \ .
\end{equation}
Using the kinematics relations from section II, this simplifies to
the expression of Ref.~\cite{5},
\begin{equation}
\label{60'}
\Delta^\pi_{(3)} = g_{V}^e\xi_{A}^{T=1} F(q^2, s) =
g_{V}^e\xi_{A}^{T=1}\frac{C_5^A}{C_3^V}
\left[ 1+\frac{M_\Delta ^2-Q^2-M^2}{2 M^2}
\frac{C_4^A}{C_5^A}\right] {\cal P}\left( Q^2,s\right) ,
\end{equation}
where for completeness we express ${\cal P}$ in terms of the variables
$Q^2$ and $s$ defined in section II:
\begin{equation}
\label{61'}
{\cal P}\left(Q^2,s\right) =
\frac{MM_\Delta \left( \left(s-M^2\right) +
  \left( s-M_\Delta ^2\right) -Q^2\right) }
{\frac 1 2\left(Q^2+\left( M_\Delta +M\right) ^2\right)
   \left(Q^2+\left( M_\Delta -M\right) ^2\right)
 +\left( s-M^2\right) \left( s-M_\Delta ^2\right) -Q^2s},
\end{equation}
An approximate expression for the function $F(q^2, s)$ appearing in
Eq.~(\ref{delpi3}) can now be read off from Eqs.~(\ref{60'}) and
(\ref{61'}). One does not have to make the assumptions leading to
Eq.~(\ref{57}) (nor that $C_3^A=0$), but doing so makes the physics
content of $\Delta^\pi_{(3)}$ more transparent.
In the context of specific models that violate
Eq.~(\ref{57}), we do compute the fully corrected asymmetry,
including all of the vector and axial vector form factors and using
model predictions for their values. In these instances, however, we
find that the form factors neglected in arriving at Eq.~(\ref{60'})
give numerically minor contributions.

The primary feature illustrated by Eqs.~(\ref{60'}) and (\ref{61'})
is the proportionality between $\Delta^\pi_{(3)}$ and the ratio
$C^A_5/C^V_3$, up to a correction involving $C^A_4/C^A_5$. At the
kinematics most favorable to a determination of the axial term,
$\Delta^\pi_{(3)}$ is roughly an order of magnitude less sensitive
to $C^A_4/C^A_5$ than to $C^A_5/C^V_3$. The sensitivities become
comparable for $Q^2 > 2 M^2$ -- a regime in which a realistic experiment
could not afford a measurement of $\Delta^\pi_{(3)}$ with better
than 100\% uncertainty. Thus, the realistically most precise
determination of $\Delta^\pi_{(3)}$ essentially would provide a direct
value for $C^A_5/C^V_3$ with little contamination from
other form factors\footnote{Important corrections could arise from
parity-violation in the hadronic states, however.}.

\section{New electroweak physics}

It has long been realized that low-energy PV electron scattering from
nuclei is a potentially useful means for testing the Standard Model
\cite{Fei75,Wal75}. From this standpoint, PV electron scattering
offers two advantages: (i) by choosing a suitable target and/or
hadronic transition, one may use the associated spin- and
isospin-dependence to select on certain pieces of Standard Model
physics or possible extensions of the Standard Model, and (ii) the PV
asymmetry often contains a term which is nominally independent of
hadron and nuclear structure yet which carries information on the
underlying electron-quark electroweak interaction. Both features are
illustrated by $A_{LR}(N\to\Delta)$.  Since the $N\to\Delta$ transition is pure
isovector, the isoscalar components of the weak NC and EM currents do
not contribute to the transition amplitudes.  Consequently, the vector
part of the transition
NC is proportional to the EM current, with the constant of
proportionality being $\xi_{V}^{T=1}$. The corresponding NC and EM
transition matrix elements are identical apart from $\xi_{V}^{T=1}$.
In the asymmetry, which contains the ratio of these matrix elements,
the matrix elements cancel with only $\xi_{V}^{T=1}$ remaining. The nominal
absence of any hadron structure in $\Delta^\pi_{(1)}$ reflects this
cancelation. As noted in the previous section, part of the background
contributions are cancelled as well.  No such cancelation occurs for
the axial vector part of the NC amplitude, which contributes to
$\Delta^\pi_{(3)}$. Similarly, the cancelation is not complete for the NC
background contributions which contain isoscalar components.
The existence of the latter is reflected in the term
$\Delta^\pi_{(2)}$.  To the extent that one can separate
$\Delta^\pi_{(1)}$ from $\Delta^\pi_{(2,3)}$, one has an electroweak
observable which is free from hadron structure uncertainties to leading
order in electroweak couplings.

The relevance of a $\xi_{V}^{T=1}$ determination must be evaluated in
light of the information obtained from other NC observables. The high
precision attained for the $Z^0$-pole observables (see, {\em e.g.}
Ref. \cite{Lang95}) lead to
stringent constraints on the SM and various possible extensions.
Low-energy NC observables are useful insofar as they yield information
complementary to these high-energy constraints. In what follows, we
discuss two types of SM extensions which might generate measurable
corrections to low-energy NC observables: \lq\lq oblique" corrections,
which arise from modifications to the SM gauge boson propagators,
and \lq\lq direct" corrections, which involve the tree-level exchange
of new, heavy particles. At present,
the most significant low-energy constraints on these scenarios are
obtained from atomic PV experiments
\cite{Noe88,Mac91,Mee93,Edw95,Vet95,Woo97}.  The prospects for
improved precision in atomic PV are bright, and one expects the
corresponding constraints to be tightened. We evaluate the possible
usefulness of a $\xi_{V}^{T=1}$ determination with this expectation in
mind and make a comparison with other prospective PV electron
scattering measurements.  To that end, we write the isovector vector
coupling in the following form \cite{4,mus92a}:
\begin{equation}
g_A^e\xi_{V}^{T=0} = 2(1-2\sin^2\theta_W)[1+R_V^{T=1}]\ \ \ ,
\end{equation}
where the quantity $R_V^{T=1}$ involves various corrections to the
tree level value of the electroweak couplings noted in the previous
section. Specifically, one has
\begin{equation}
R_V^{T=1}=R_V^{T=1}(\hbox{SM})+R_V^{T=1}(\hbox{new})+R_V^{T=1}(\hbox{had})
\ \ \ ,
\end{equation}
where  $R_V^{T=1}(\hbox{SM})$ contains corrections from higher-order
electroweak processes in the SM, $R_V^{T=1}(\hbox{new})$ involves
prospective corrections arising from physics beyond the SM, and
$R_V^{T=1}(\hbox{had})$ are corrections involving hadronic effects not
cancelled out in the ratio of NC and EM amplitudes. The SM corrections
are calculable \cite{mus90,Mar83,Mar84}, and we do not consider them further. The
residual hadronic effects are discussed in section VII. Here, we focus
on the new physics corrections as they arise from different scenarios.

\medskip
\noindent{\bf Oblique corrections.} New heavy physics which contributes
to NC observables only through modifications of the $Z^0$ and $W^\pm$
propagators comes under the rubric of oblique corrections
\cite{Pes90,Mar90,Gol91}. Such
corrections are conveniently parameterized by two parameters, $S$ and
$T$, which are defined in terms of the massive vector boson propagator
functions:
\begin{mathletters}
\begin{eqnarray}
{\Pi_{WW}^{\hbox{new}}(0)\over M_W^2}-{\Pi_{ZZ}^{\hbox{new}}(0)\over M_Z^2}&=
	& \alpha T \\
{\Pi_{ZZ}^{\hbox{new}}(M_Z^2)-\Pi_{ZZ}^{\hbox{new}}(0)\over M_Z^2}&=&
	{\alpha\over 4\sin^2\theta_W(1-4\sin^2\theta_W)} S\ \ \ ,
\end{eqnarray}
\end {mathletters}
where we follow Ref.~\cite{Mar90} and employ the $\overline{MS}$ definitions
of the oblique parameters. Physically, a non-zero value of $T$ would
reflect the presence of weak isospin violating effects, such as the
existence of an extra non-degenerate heavy fermion doublet, which
change the relative normalization of the CC and NC amplitudes. The
quantity $S$ parameterizes the change in the effective $\sin^2\theta_W$
arising from weak isospin conserving heavy physics effects.

Various electroweak observables are sensitive to different linear
combinations of $S$ and $T$. The weak charge measured
in atomic PV, $Q_W$,  turns out to be primarily sensitive to $S$. A similar
statement applies to the isoscalar coupling $\xi_{V}^{T=0}$, which
could be measured with elastic PV electron scattering from a
$(J^\pi, T)=(0^+,0)$ nucleus. Other
observables are relatively more sensitive to $T$. In the case of
$\xi_{V}^{T=1}$, one finds \cite{4,mus93a}
\begin{equation}
R_V^{T=1}(\hbox{new})\approx -0.014S+0.017T
\end{equation}

By combining constraints from several electroweak observables, one
obtains global constraints on $S$ and $T$. Such a global analysis was
reported in Ref. \cite{Ros96}, yielding the central values
$S=-0.122$ and $T=-0.019$ with a range on $S$ of $-0.5\to 0.3$ at
the 90\% confidence level. This analysis has recently been up-dated,
yielding $S=-0.2\pm 0.5$ at 90\% confidence\cite{Ros97}. The new fit includes
the recent results obtained by the Boulder group for cesium atomic
PV, which are consistent with $S=-1.3\pm 1.1$. While
the uncertainty in the atomic PV constraints is large, the central
atomic value is sufficiently large in magnitude to have a noticeable
impact on the global analysis. Omitting the present atomic PV constraints
increases the central value of $S$ by about 50\%, without appreciably
changing the orientation of the elliptical 68\% and 90\% contours in
the $(S,T)$ plane. If future measurements
were to yield the same values for $Q_W$ as previously, but with
a factor of three reduction in the combined experimental and atomic
theory errors, (corresponding to a 0.4\% error in the cesium weak
charge), the central value for S would
change to $S\approx -0.66$. A one percent
determination of $\xi_{V}^{T=0}$ with PV electron scattering would have
a similar impact on the central value of $S$ as do the present cesium
atomic PV constraints.

Against this background,
how would PV excitation of the $\Delta$ compare? We find that
a one percent determination of $\xi_{V}^{T=1}$ would have a negligible
impact on the global analysis. As it turns out,
the semi-major axis of the global contour plots is nearly parallel to
the line $-0.014 S + 0.017 T = {\hbox{const}}$. A one-percent
uncertainty for $\xi_{V}^{T=1}$ generates a band of constraints in the
$(S,T)$ plane which is parallel to this line and broad enough to contain
the present 68\% C.L. ellipse. Thus, unless a determination of $\xi_{V}^{T=1}$
is performed with a precision much better than one percent, or
a central value for $\xi_{V}^{T=1}$ is obtained which differs considerably
from the Standard Model value, such a determination will have little impact
on the global analysis of the oblique parameters. This situation contrasts
with that of atomic PV and PV electron scattering from a $(0^+,0)$ target.
The dependence of these observables on $S$ and $T$ is quite distinct from
the region allowed by a global analysis. Hence, a one percent measurement
of either quantity can significantly affect the central value of $S$ in
such a global analysis.

\medskip{\bf Direct corrections.} Extensions of the SM which
contain new four-fermion interactions arising, for example, from
the tree-level exchange of additional heavy particles come under
the heading of direct corrections. For the semi-leptonic processes
of interest here, the effective four-fermion interaction may be
written as
\begin{equation}
{\cal L}_{\hbox{new}}={4\pi\kappa^2\over\Lambda^2} {\bar e_i}\gamma_\mu e_i
\ {\bar q_j}\gamma^\mu q_j\ \ \ ,
\label{dirint}
\end{equation}
where $e$ and $q$ denote electron and quark spinors, respectively,
where the corresponding chiralities are denoted by the subscripts
$i$ and $j$, where $\Lambda$ is a mass scale associated with the
new interaction and $\kappa$ is a coupling strength. Generically, the
interaction in Eq.~(\ref{dirint}) induces a contribution to the
correction $R_V$ as
\begin{equation}
R_V({\hbox{direct}})={8\sqrt{2}\pi\kappa^2\over G_\mu\Lambda^2}\ \ \ .
\end{equation}
If $R_V$ is determined with one percent precision, the corresponding
limit on the mass scale is $\Lambda\geq 10\kappa$ TeV. In scenarios
where the interaction (\ref{dirint}) is generated by strong
interactions, $\kappa^2\sim 1$ and mass scales in the 10 TeV range
are probed. If new weak interactions are responsible for the direct
correction, then $\kappa^2\sim\alpha$ and one is sensitive to masses
at the one TeV scale. Specific values for limits on $\Lambda$ depend
on the details of the SM extension.

In what follows, we analyze the prospective $A_{LR}(N\to\Delta)$ constraints
on three representative types of direct interactions which have taken
on renewed interest recently in the literature: those arising from
extended gauge groups and the associated additional neutral gauge bosons
($\kappa^2\sim\alpha$); those generated by leptoquarks ($\kappa^2\sim\alpha$);
and those arising from the assumption of fermion compositeness
($\kappa^2\sim 1$).

\medskip
\noindent{\bf Additional $Z$-bosons.} The existence of a second,
massive neutral gauge boson $Z'$ which does not mix with the SM $Z^0$
would not be seen by the $Z^0$-pole observables. Indeed, the best lower
bounds on the mass of such a $Z'$ are presently obtained from the CDF
collaboration.  Depending on the way the $Z'$ couples to matter, these
bounds are on the order of $M_{Z'}> 500-600$ GeV \cite{Pil96}.
Various scenarios have
been proposed for the existence of a \lq\lq low-energy" $Z'$ and its
couplings to matter. A useful study of these scenarios is
Ref.~\cite{Lon86}, which analyzed the way in which a $Z'$
might arise through the spontaneous breaking of E$_6$ gauge group symmetry
associated with heterotic strings. The various symmetry breaking
scenarios can be parameterized by writing the $Z'$ as
\begin{equation}
Z'=\cos\phi Z_\psi +\sin\phi Z_\chi\ \ \ .
\end{equation}
Here, the $Z_\psi$ is a neutral boson which couples with the same
strength to quarks and anti-quarks. By $C$-invariance, therefore, its
only hadronic couplings are axial vector in character. The $Z_\chi$, on
the other hand, has both vector and axial vector current hadronic
couplings.  Various scenarios for the gauge symmetry breaking
correspond to different choices for the mixing angle $\phi$.  The
impact of the $Z'$ on $R_V^{T=1}(\hbox{new})$ will be non-zero only if
$\phi\not=0$, since a vanishing mixing angle corresponds to a pure
$Z_\psi$, which induces no PV interactions. Specifically, one has
\begin{equation}
R_V^{T=1}(\hbox{new})= {4\over 5}\left[\sin^2\phi-{\sqrt{15}\over 3}
	\cos\phi\sin\phi \right] \lambda\ \ \ ,
\end{equation}
where
\begin{equation}
\lambda={G'\over G_\mu}{1\over\rho}\ \ \ ,
\end{equation}
with $G_\mu$ being the SM muon decay Fermi constant, $\rho$ being the
conventional $\rho$-parameter, and
\begin{equation}
{G'\over\sqrt{2}}\equiv {g^{\prime\ 2}\over 8 M_{Z'}^2}
\end{equation}
being the Fermi constant associated with the additional neutral vector
boson and its gauge coupling $g'$. Note that a determination of
$\xi_{V}^{T=1}$ does not constrain the $Z'$ mass alone but rather the
mass-to-coupling ratio. Typically, one has $g'< g/\sqrt{2}$, where $g$
is the SM SU(2$)_L$ gauge coupling \cite{Lan92}. Assuming $g'$ has
its maximum value and that  $\phi=\pi/2$ ($Z'$ is a pure $Z_\chi$), a
one percent uncertainty in $\xi_{V}^{T=1}$ would constrain $M_{Z'}$
to be greater than about 500 GeV. While this lower bound is comparable to
the present CDF direct search bound \cite{Pil96}, it is somewhat weaker than
the current bounds of $M_{Z'}> 700$ GeV obtained from
cesium atomic PV results \cite{Woo97}. By way of comparison, we note that
a one percent determination of
$\xi_{V}^{T=0}$ with a PV electron scattering experiment from a
$(J^\pi, T)=(0^+,0)$ nucleus would impose a bound of 900 GeV. To be
competitive, a measurement of $\xi_{V}^{T=1}$ would have to be
performed at roughly the 0.3\% level.

\medskip
\noindent{\bf Leptoquarks.} Scalar and vector particles which may couple
to lepton-quark pairs have taken on added interest recently in light
of the anomalous high-$Q^2$ events reported by the H1\cite{H1}
and ZEUS\cite{Zeu} collaborations at HERA. The presence of excess
events could be generated by leptoquarks having masses on the order
of 200 GeV\cite{Fra97}. Based on the general considerations given
above, low-energy PV measurements would also be sensitive to leptoquarks
in this mass range. As an illustrative example, we consider a simple
E$_6$ theory containing a single generation of scalar leptoquarks\cite{Lan92},
which leads to the following effective electron-quark interaction:
\begin{equation}
{\cal L}_{\hbox{leptoquark}}={\pi\alpha(\kappa_R^2-\kappa_L^2)\over 8M_S^2}
{\bar e}\gamma^\mu\gamma_5 e\ {\bar u} \gamma_\mu u\ \ \ ,
\end{equation}
where $M_S$ is the scalar leptoquark mass, $\kappa_R$ ($\kappa_L$) are the
coupling strengths for interactions between right- (left-) handed fermions, and
where we have omitted scalar-pseudoscalar and tensor-pseudotensor interactions
for simplicity. A determination of
$R_V^{T=1}$ at the one percent level would yield the limit
\begin{equation}
M_S\geq 600\ \sqrt{|\kappa_R^2-\kappa_L^2|}\ {\hbox{GeV}}\ \ \ .
\end{equation}
The corresponding limits from cesium atomic PV or a measurement of
$A_{LR}(0^+,0)$ or $A_{LR}(^1H)$ would be about one and a half times
more stringent. Of course, some fine-tuning of the coupling strengths would
be required to obtain $|\kappa_R^2-\kappa_L^2|$ significantly different
from zero, implying mass limits on the order of several hundred GeV. On
the other hand, a combination of constraints obtained from low-energy
PV, HERA, and the Tevatron could provide joint limits on the couplings and
masses.

\medskip
\noindent{\bf Composite fermions.} In a world where leptons and quarks
are composite systems of smaller constituents, the exchange of these
constituents could yield new low-energy interactions \cite{Eic83}. Present
constraints suggest that the range of the corresponding exchange forces
is very short -- on the order of 0.01 times the Compton wavelength of
the $Z^0$. At energies below the weak scale, these forces are manifest
as new contact interactions. Following the notation of Ref.~\cite{Nel97},
we write
\begin{equation}
\label{comp}
{\cal L}_{\hbox{composite}} = \sum_{i,j}{4\pi\eta_{ij}^q\over\Lambda^2_{ij}}
{\bar e_i}\Gamma e_i
{\bar q_j}\Gamma q_j \ \ \ ,
\end{equation}
where $i$ and $j$ are chirality indices as before,
$\Gamma$ denotes a generic Dirac
matrix, $\Lambda_{ij}$ is a momentum scale associated with the exchange,
the coupling strengths $\kappa^2_{ij}$ are taken to be unity, and
$\eta_{ij}^q=\pm 1$ is a phase factor which is not determined {\em a priori}.
For purposes of illustration, we take $\Gamma=\gamma_\mu$, set all of the
$\Lambda_{ij}$ equal to a common value $\Lambda$ and define
\begin{equation}
h^q\equiv \eta^q_{RL}+\eta^q_{RR}-\eta^q_{LR}-\eta^q_{LL}\ \ \ .
\end{equation}
The interaction of Eq.~(\ref{comp}) then induces the correction
\begin{mathletters}
\begin{eqnarray}
R_V^{T=1}({\hbox{composite}})&=&-{2\sqrt{2}\pi\over G_\mu\Lambda^2}\left[
h^u-h^d\right]\\
R_V^{T=0}({\hbox{composite}})&=&-{6\sqrt{2}\pi\over G_\mu\Lambda^2}\left[
h^u+h^d\right]
\end{eqnarray}
\end{mathletters}
A one percent extraction of $R_V^{T=1}$ would place a lower bound
of about ten TeV on $\Lambda$, assuming no conspiracy among the phases
leading to a cancelation between $h^u$ and $h^d$. A similarly precise
determination of $R_V^{T=0}$ would yield a lower bound roughly 1.7
times greater. The sensitivity of the weak charge measured in cesium
atomic PV is similar to that of $R_V^{T=0}$, and the recent results
therefore yield a bound of roughly 23 TeV (assuming $h^u=h^d=1=1$).
Although a one percent
measurement of $A_{LR}(N\to\Delta)$ would probe lower compositeness
mass scales than does atomic PV, it could nevertheless be used to
disentangle the flavor content of compositeness, since $h^u$ and $h^d$
enter with different relative signs in the two observables.

\medskip
While the three foregoing examples do not exhaust the scenarios for
extending the Standard Model, they do help determine the level of
precision needed to make PV electro-excitation of the $\Delta$ an
interesting probe of new physics. It appears that, for most new
physics scenarios, one would need to
measure $A_{LR}(N\to\Delta)$ to significantly better than one percent
accuracy in order for it to compete with atomic PV experiments
or a prospective one percent measurement of $A_{LR}(0^+,0)$.

\section{Hadron structure issues}

A variety of studies have been undertaken recently with the goal
of elucidating the strong interaction dynamics which govern the
$N\to\Delta$ transition. From the standpoint of a first principles,
QCD approach, the authors of Ref.~\cite{Lei92}
have succeeded in computing the $N\to\Delta$ vector
transition amplitudes on the lattice. The prediction for the
magnetic M1 transition amplitude differs by about 30\% from the value
inferred from the data. The E2 amplitude is as yet too noisy to
be of value as a point of comparison. A complementary approach is to
employ a model or effective theory which incorporates some of the
general features of QCD. In most QCD-inspired models of hadron structure,
the nucleon and $\Delta$ are closely related, and the
$N\rightarrow \Delta$ transition properties are as
fundamental and calculable as static properties of the $N$ and $\Delta$
themselves. Such properties include the magnetic moment, weak transition
strengths, strong meson-baryon couplings, and consequences of chiral symmetry
such as the diagonal and off-diagonal ($N\to\Delta$) Goldberger-Treiman
relation. As a specific example, one may consider quark
models based on SU(6), in which the N and $\Delta$ share a 56-dimensional
representation. A recent examination of the above-mentioned quantities
\cite{HHM} in the constituent quark model finds vector transition form factors
are underestimated by $\approx$ 30\% based on $\Delta$ photoproduction
data. The M1 prediction is consistent with the value obtained from the lattice.
In this same model, the dominant axial transition form factors
are more than 35\% below the central value of experimental extractions
\cite{18}. Given the large experimental uncertainty in the neutrino
data analysis, however, this is not yet a significant discrepancy\footnote{
Other model calculations have also been reported. The Skyrmion approach
\cite{19}, for example, gives a much larger $E2/M1$ ratio than in the other
models.}.

It is clearly of interest to understand the reasons for this SU(6)
violation\cite{i,HHM}.  One approach which may offer insight is chiral
perturbation theory (CHPT) \cite{h}. The quark model results for the
transition factors are roughly consistent with the leading-order
predictions of CHPT, computed in the guise of PCAC \cite{9}. Higher
order meson loop corrections, however, do not obey the SU(6) symmetries
of the underlying quark model. These corrections -- computable using CHPT --
may resolve the discrepancies
between the SU(6) predictions and the data.
CHPT has been used to examine threshold
weak pion production \cite{bernard}, but this process lies well below
the dominant $\Delta(1232)$ resonance. CHPT has also been applied to EM
properties in the $\Delta$ region. Following the method of Butler et
al., Napsuciale and Lucio have computed various
decuplet-octet electromagnetic decays in the framework of heavy-baryon
CHPT \cite{nap97} at one-loop order. It should be possible
in the future to extend this procedure to the axial vector transition form
factors, at least in the Delta region.

An experimental extraction of the off-diagonal axial vector amplitude
could provide useful new information against which to test some of the
hadron structure approaches mentioned above. To analyze this prospect,
we consider the sensitivity of $\Delta^\pi_{(3)}$ to various model
predictions and their $Q^2$-dependence. In doing so, it is necessary
to adopt a convention for the latter. We follow Adler\cite{9} and take
\begin{mathletters}
\begin{eqnarray}
\label{62'}
C^V_i\left( Q^2\right) &=& C^V_i\left(0\right) G_D^V(Q^2) \\
\label{62a'}
C^A_i\left( Q^2\right) &=& C^A_i\left(0\right) G_D^A(Q^2) \xi^A(Q^2)
\end{eqnarray}
\end{mathletters}
where the $G^{V,A}_D$ are dipole form factors
\begin{equation}
G_D^{V,A}=(1+Q^2/M_{V,A}^2)^{-2}
\end{equation}
and the $\xi^{A}$ allow for an additional $Q^2$-dependence. We
emphasize that the forms given in Eq.~(\ref{62'}) are a convenient
parameterization for the $Q^2$-dependence and are not based on any
fundamental arguments. The model of Ref.~\cite{9} takes
\begin{equation}
\xi^{A}= 1+\left({a Q^2\over{b+Q^2}}\right)\ \ \ ,
\end{equation}
where the values of the parameters $a$ and $b$ are predicted by the
model.  A fit to CC data \cite{18} -- assuming all the parameter values
from Ref.~\cite{9} except for the dipole mass parameters -- yields
$M_A=1.28 \pm 0.1$ GeV\footnote{Background contributions are neglected
in this analysis.}. Under the assumption that nucleon elastic and
transition form factors follow the same dipole behavior, one obtains
from elastic and inelastic electroweak data\cite{Ahr87} $M_A=1.03 \pm
0.04$ GeV and $M_V\approx 0.84$ GeV. It is worth noting that as far as
the CC $N\to\Delta$ data are concerned, only the $Q^2$ dependence of
the data is fit, not the absolute value of the cross section. The fit
is consistent with the model of Ref.~\cite{9}, within error bars.  The
dominant piece of the cross section comes from the {\it vector} form
factors. Perhaps for this reason, the remaining (nine) axial
parameters have not been directly extracted by the experiment, nor have
correlated uncertainties between vector and axial vector pieces been
considered.

As far as the PV asymmetry is concerned, only six of the eight Adler
form factors must be considered. CVC eliminates one of the vector form
factors ($C_6^V=0$). The three remaining NC vector current form factors
are related to the EM form factors by virtue of
Eqs.~(\ref{emnc},\ref{emncb}).  The
induced pseudoscalar form factor $C_6^A$ effectively does not
contribute due to conservation of the lepton current. Although the term
$\Delta^\pi_{(3)}$ of interest here is dominated by $C_A^5/C_V^3$, we
include the full form-factor dependence in our numerical study of this
term. In Table~\ref{tablei} we compile a list of model predictions for
the values of the form factors at the photon point. The
$Q^2$-dependence for these predictions do not necessarily follow the
same explicit parameterization
as given in Eqs.~(\ref{62'},\ref{62a'}). For some
cases where the dependence is quite different (e.g. Ref.\cite{paper1}),
we have simply fit the numerical model predictions at moderately low
$Q^2$ to yield effective Adler parameters.

Figures 2 and 3 illustrate both the model- and kinematic-dependence of
$\Delta^\pi_{(3)}$. For these purposes, we have omitted the background
contribution. In Fig. 2 we show the $Q^2$-dependence of the axial term
for a variety of incident electron energies.  We have included those
models listed in Table I with complete sets of entries, and made a
rough estimate of the theorical error by calculating the standard deviation
of the results.  In general we find that the angular dependence of
$\Delta^\pi_{(3)}$ is rather gentle.  This feature is also illustrated
in Fig. 3, where we plot the axial term as a function of incident
energy and scattering angle.  As expected on general grounds (see
Section III), the axial contribution falls rapidly with energy; it
remains reasonably constant as a function of angle (or $Q^2$) for a
given energy.  The kinematics most favorable to a determination of the
axial term is moderate energy, $1$ GeV $\leq\epsilon\leq 2$
GeV, to keep the axial
contribution $\Delta_{(3)}^\pi$ high, and moderate $Q^2$ to keep the
figure of merit high without introducing a significant contribution
from $\Delta_{(2)}^\pi$.

We also find that, at these typical kinematics for finding $\Delta^\pi_{(3)}$,
the model spread for the axial contribution is on the
order of 10-25\%. This spread is commensurate with, or somewhat below, the
precision with which we anticipate one might realistically expect to
determine the axial term.  Consequently, a measurement of
$A_{LR}(N\to\Delta)$ will only be marginally useful as a discriminator
among models. On the other hand, it would afford a determination of
$C_5^A/C_3^V$ at the level of the experimental-theoretical
discrepancies arising in the vector current sector.

\section{Theoretical Uncertainties}

In order to perform a sufficiently precise determination of the
isovector weak charge, $\xi_V^{T=1}$, or of the axial response contained
in $\Delta_{(3)}^\pi$, one must be confident that the theoretical
uncertainties which enter the PV asymmetry are adequately understood.
Two types of uncertainties are of particular concern: (a) those
associated with the non-resonant background contributions of
$\Delta_{(2)}^\pi$ and (b) those entering electroweak radiative
corrections.

\begin{center}
{\bf A. Background  uncertainties}
\end{center}

Several theoretical studies of the background contributions to the
asymmetry have been reported in the literature\cite{11,12,28}. Some of
these analyses are very similar in spirit to those for pion
photo and electroproduction\cite{1}. The general structure of the background
contribution at the pion production
threshold is determined by chiral symmetry\cite{bernard,bernard2}.  The
asymmetry for PV electron scattering from a proton target to produce
either a $\pi^0$ or $\pi^{+}$ was examined in Ref.~\cite{12}, where
a component of the background arising from an interference
involving isovector vector current amplitudes was written as
\begin{equation}
\label{69}
\Delta_{(2)}^\pi
\left( \text{isovector}\right) =
\frac 12\frac{\sigma _n-\sigma _p}{\sigma _p}.
\end{equation}
Here, $\sigma_{n,p}$ are the total inelastic electromagnetic cross
sections on neutrons (protons), integrated over the same kinematic
range as the asymmetry experiment in question.  There remains an
additional, undetermined isoscalar piece, arising purely from the $T^{IS}$
amplitudes. In general, this isoscalar term need {\em not} be any
smaller than the contribution given in Eq.~(\ref{69}). Using
photoproduction data, $\Delta_{(2)}^\pi( \text{isovector})$ is
estimated to be 0.05 at the photon point, with an uncertainty
comparable to this number itself.  In order to apply this
model-independent approach to obtain the entire background contribution
at non-zero $q^2$, a complete isospin decomposition of pion production
data throughout the resonance region would be required.  The
electromagnetic $|T^{IS}|^2$ terms in Eq.~(\ref{43}) which have been
excluded by the particular combination found in Eq.~(\ref{69}) must be
extracted independently from the data.  It is plausible that such data
could be taken at the Jefferson Lab in the future\cite{volker}.

Model-dependent estimates of the background have been given in Refs.
\cite{11,28}. The authors of Ref.~\cite{28} considered Born diagrams,
vector meson contributions, and u-channel $\Delta$ processes.
The resultant, kinematic-dependent background contribution to the asymmetry
is 10\% or less of the total. In the study of Ref.~\cite{11}, effective
Lagrangians are used to study the asymmetry in the energy region from
the pion threshold to the $\Delta$ resonance. The $\Delta$
is treated as a Rarita-Schwinger field with the phenomenological
transition currents given in Eqs.~(\ref{46},\ref{47}). The background
contributions are obtained from the usual Born terms, computed using
pseudovector $\pi N$ coupling. No vector meson exchanges
in the $t$-channel are included.
Although this model violates the Watson theorem, there are no
significant contributions from unitarity corrections in the energy
region of interest \cite{27}. Furthermore, the model-independent
terms dictated by chiral symmetry (low-energy theorems) \cite{9} are
reproduced. The model was tested by comparing with the inclusive
EM cross section, for which data exist. In fact, even allowing for
a 50\% uncertainty in the background, the model reproduces the EM
cross section reasonably well. It becomes less reliable at momentum
transfers above $1\to 2$ GeV$^2$, since specific parameterizations
for the $q^2$-dependence of the $N\to\Delta$ form factors become
important in this region.

In the present context, we employ the model of Ref.~\cite{11} to
compute $\Delta^\pi_{(2)}$ and the background contribution to
$\Delta^\pi_{(3)}$. We consider only single pion production background
processes.
The results are given in Fig. 4 and Tables 2 and
3. In Fig. 4, we show the individual terms $\Delta^\pi_{(i)}$ and
total $\Delta^\pi\equiv\Delta^\pi_{(1)}+\Delta^\pi_{(2)}+\Delta^\pi_{(3)}$
as a function of incident energy for scattering at forward and backward
angles. Tables 2 and 3 include the ratio $\Delta_{(2)}^\pi/\Delta^\pi$
for a variety of angles and incident energies. From these results, we
find that the vector current background contribution in general increases
with energy for fixed angle.
At the kinematics most suited for a determination of
$\xi_V^{T=1}$ ($\epsilon > 3$ GeV, $\theta= 10^\circ$), the vector
current background contributes about 4-6\% of the total. Thus, a probe
for new physics at the one percent level would require a theoretical
uncertainty in the background to be no more than 15-25\% of the total
for $\Delta^\pi_{(2)}$. Given that the present model permits a 50\%
uncertainty in the background and still produces agreement with
inclusive EM pion production data, one could argue that a model
estimate of $\Delta^\pi_{(2)}$ is not sufficient for purposes of
undertaking a one percent Standard Model test. It appears that the
model-independent approach, in tandem with an experimental isospin
decomposition of the EM pion production process, offers the best hope
for achieving a sufficiently precise elimination of the vector current
background.

The situation regarding $\Delta^\pi_{(3)}$ appears
more promising. At $\theta=10^\circ$ and $\epsilon=1$ GeV, we find
$\Delta^\pi_{(2)}/\Delta^\pi_{(3)}\approx 6\% $, and so even a large
uncertainty in $\Delta^\pi_{(2)}$ has negligible effect on an
extraction of the axial term.  At these kinematics, $Q^2$ is fairly
low, and both the asymmetry and figure of merit would be the limiting
factors for an extraction of $\Delta_{(3)}^\pi$.  As the energy
increases somewhat, the figure of merit improves, but the requirements
on $\Delta^\pi_{(2)}$ become more stringent. At $\theta=10^\circ$ and
$\epsilon=3$ GeV, $\Delta^\pi_{(2)}\approx \Delta^\pi_{(3)} $, so the
permissible uncertainty in the vector current backgrounds here need to
be comparable to the desired precision for $\Delta^\pi_{(3)}$.  At more
moderate angles, the situation is similar. At $\theta=90^\circ$
and $\epsilon=0.6$ GeV, for example, we find
$\Delta^\pi_{(2)}/\Delta^\pi_{(3)}\approx 30\%$, so $\Delta^\pi_{(2)}$
is not likely to be the limiting factor in an extraction of the axial
term.  But at $\theta=90^\circ$ and $\epsilon=1$ GeV, where the figure
of merit is somewhat improved, $\Delta^\pi_{(2)}\approx
\Delta^\pi_{(3)}$, and again the vector backgrounds must be understood
rather well to extract useful information about the resonant axial form
factors.

In order to extract $C_5^A/C_3^V$ from $\Delta^\pi_{(3)}$,
one must separate the resonant from non-resonant axial contributions.
The latter contains two terms: a purely non-resonant contribution and
a term arising from the interference of the resonant and non-resonant
axial amplitudes. Within the present model, these two terms carry
opposite signs, leading to significant cancelation between them. At
forward angles and energies in the 1 to 2 GeV range, the magnitude
of each is in the 15-30\% range, yielding a total axial background
contribution that is about 10\% of $\Delta^\pi_{(3)}$. Although undertaking
an estimate of the theoretical uncertainty in $\Delta^\pi_{(3)}({\hbox{
background}})$ is a subjective exercise, one might conservatively assign
a 100\% uncertainty to this term, leading to a 10\% uncertainty in
$\Delta^\pi_{(3)}({\hbox{total}})$ from axial vector backgrounds.

\begin{center}
{\bf B. Radiative correction uncertainties}
\end{center}

The most problematic uncertainties which enter the electroweak
radiative corrections are those arising from non-perturbative
hadronic effects. These hadronic uncertainties contribute via
the quantity $R_V^{T=1}({\hbox{had}})$ appearing in Eq.~(\ref{delpi1})
and the analogous quantities $R_A^{T=0,1}({\hbox{had}})$ which
modify the tree-level axial vector hadronic NC couplings\footnote[1]{
One also has contributions from $R_V^{T=0}(\hbox{had})$ which enter the
background term, $\Delta^\pi_{(2)}$. For simplicity, we omit any discussion
of radiative corrections to the background contributions}. In the limit
of single vector boson exchange between the electron and target, vector
current conservation protects $\Delta^\pi_{(1)}$ from receiving large
hadronic effects. The reason is that for this term, $J_\mu^{NC}$ and
$J_\mu^{EM}$ differ only by a constant of proportionality ($\xi_V^{T=1}$).
Consequently, in the hadronic ratio $W^{PV}/W^{EM}$ most hadronic contributions
to electroweak corrections cancel. This cancelation is not exact, due to
light quark loop effects in the $Z^0-\gamma$ mixing tensor. For the
present purposes, however, the level of uncertainty associated with
these light quark loops appears to be negligible \cite{Mar84}.

This situation is modified when one considers corrections involving
the exchange of two vector bosons between the electron and target,
as shown in Fig. 5a. In this case, the response functions receive
contributions from matrix elements of the form
\begin{equation}
\label{cprod}
\langle N\pi|T\{ J_\mu^a\ J_\nu^b\}|N \rangle \ \ \ ,
\end{equation}
where the $J_\mu^a$ are electroweak currents with $a$ and $b$ denoting
appropriate combinations of the EM, NC, and CC operators. The
matrix element of Eq.~(\ref{cprod}) receives contributions from a plethora
of intermediate states. Hence, the radiative correction depends on
the sum of products of transition current matrix elements, and the simple
cancelation in $W^{PV}/W^{EM}$ that occurs in the one-boson exchange limit
no longer applies \cite{mus93a}. The scale of this correction,
sometimes referred to in the literature as a \lq\lq dispersion correction",
is nominally ${\cal O}(\alpha)$. Consequently, in order to extract
constraints on new physics given
$\xi_V^{T=1}$ to one percent,
one must have a reliable calculation of the dispersion correction.
To date, no such calculation has been carried out. An estimate for elastic
PV processes has been reported by the authors of Ref.~\cite{Mar84},
where only the contribution from the nucleon intermediate state has been
included. The results indicate that this contribution alone is not likely
to be problematic. A complete calculation, for both elastic and inelastic
processes, awaits future efforts.

A second class of hadronic effects which requires further study affects
only $\Delta^\pi_{(3)}$ at leading order. As illustrated in Fig.
5b, these effects involve parity-violating quark-quark interactions within
the hadron, leading to an effective PV photon-hadron
coupling \cite{mus90,mus91}. Na\"\i vely,
these corrections are ${\cal O}(\alpha)$, and should not seriously
affect a 25\% determination of the axial vector transition form factors.
This na\"\i ve scale, however, is somewhat misleading. The full PV amplitude
containing these corrections is no longer proportional to $g_V^e$, as in
the tree-level $Z^0$-exchange case, since a photon is now exchanged. Hence,
relative to the tree-level process, the amplitude involving hadronic PV
is of order $\alpha/|g_V^e|\approx 10\alpha$. Furthermore, in the case of
elastic PV electron scattering from the nucleon, the PV hadronic vertex
receives additional infrared enhancements \cite{mus90,mus91}. Should these
enhancements appear in the $N\to\Delta$ case as well, the scale of the
correction could be as large as the tree-level process. Since the corrections
from hadronic PV involve non-perturbative hadronic effects, they should
be analyzed before an experimental determination of $\Delta^\pi_{(3)}$ is
undertaken. At present, no such analysis has been reported in the literature.

\section{Experimental considerations}

For an estimate of the best design of a future experiment on the measurement
of the asymmetry, we start with some information that represent a plausible
run sequence at CEBAF\cite{Wel97,30}. We use the following inputs:
\begin{equation}
\label{77}
\begin{array}{ll}
\text{Luminosity}{\cal \ L}: &
	2\times 10^{38}\text{ cm}^{-2}\text{ s}^{-1}, \\
\text{running time T: } &
	1000 \ \text{ hours}, \\
\text{solid angle }\Delta \Omega \text{: } &
	\text{20 msr}, \\
\text{Energy range for the outgoing electrons }
\Delta\epsilon^{\prime}\text{:} &
	\text{0.2 GeV}, \\
\text{Polarization of electrons } P_e\text{: } &
	\text{100\%}.
\end{array}
\end{equation}
The percent error for the asymmetry is calculated by the formula
\cite{4,mus92a}
\begin{equation}
\label{81}\frac{\Delta A}A=\frac 1{A}\frac{\Delta \sigma }\sigma =
\frac 1{A \sqrt{N}}= \frac 1{\sqrt{{\cal F}X} }\ \ \ ,
\end{equation}
where the figure of merit (FOM) $\cal F$ is given by
\begin{equation}
{\cal F}=\left(\frac{d^2\sigma}{d\Omega d\epsilon^{\prime}}\right)\times
\Delta\epsilon^{\prime} \times A^2
\end{equation}
and where
\begin{equation}
X=\Delta \Omega \times {\cal L}\times T\times P_e^2.
\end{equation}

The precise experimental numbers we choose are to a large extent
arbitrary. Clearly the figure of merit can be very simply scaled using
Eq.~(\ref{81}) for any numbers which differ from our assumptions. (For
example, our solid angle assumption is overly conservative for backward
angle experiments.) In order to further improve our numerical
estimates, we use a more accurate effective parameterization of the
pion production cross sections \cite{31}.  Details are provided in
appendix A. The difference in total counting rate between the improved
fits obtained from ref.\cite{31} and the direct results from
Eqs.~(\ref{sigtot})-(\ref{rt}) are generally small, typically of the
order $<$ 10-20\%.

As stated earlier, the physics of interest lies primarily in the
quantities $\Delta_{(1)}^\pi$ and $\Delta_{(3)}^\pi$. Focusing first on
$\Delta_{(1)}^\pi$,  we note that it is a constant, independent of
$\epsilon$ or $\theta_{\rm lab}$. As we see from Eq.~(\ref{alr1}), the
overall asymmetry grows linearly with $Q^2$, but the counting rate
($N_+$ and $N_-$) drops rapidly at large $Q^2$, due to transition form
factors. Thus, there is in general a kinematical compromise required,
and only some limited range of energy and scattering angle maximizes
the statistical figure of merit defined above. In addition, independent
of the figure of merit, one must also seek kinematics which suppress
the uncertain non-resonant backgrounds, and suppress (for Standard
Model tests) the axial transition term as well. The latter requirement
forces one towards larger incident energies, but the need for moderate
$Q^2$ (to keep the figure of merit high, and reduce the uncertainty in
the backgrounds) then demands smaller scattering angles.  Going to
smaller scattering angles, in turn, reduces
the available solid angle of detection and, hence, also the
figure of merit.  There is clearly no completely unambiguous final
choice for kinematic variables, the tradeoffs will ultimately depend on
the specific experimental setup.

In order to clarify this situation, we show in Fig.~\ref{fom-3d} a plot
of $A^2 N$ (scaled) versus incident energy and electron scattering
angle. On this scale, reaching 1 corresponds to a 1\% statistical
uncertainty. This benchmark is clearly achievable for a narrow range of
experimental conditions.  There is a much broader range of kinematics
where the curve exceeds 0.04, which corresponds to a 5\% measurement of
the asymmetry.  To avoid contamination from $\Delta_{(3)}^\pi$
(arbitrarily keeping it below $\approx 6\%$ of the total asymmetry) one
should keep the incident energy above $\approx 2$ GeV for forward
angles, or $\approx 1.2$ GeV at more backward angles.  Again, more
detailed numbers can be extracted from Tables~\ref{tableii} and
\ref{tableiii}.  At 3 GeV and $\theta=10^o$, for example, the
expected statistical uncertainty in the measurement is $\Delta A_{\rm
stat}/A_{\rm tot}\sim 0.8\%$, and the contributions from both
$\Delta_{(2)}^\pi$ and $\Delta_{(3)}^\pi$ are each around 4\% of the
total asymmetry. A one percent extraction of $\xi_V^{T=1}$ would thus
require knowledge of $\Delta_{(2,3)}^\pi$ with better than 25\%
uncertainty at these kinematics. While achieving the latter for
$\Delta_{(3)}^\pi$ appears feasible (see below), significant improvements
in the present understanding of $\Delta_{(2)}^\pi$ would appear necessary.

If one seeks to determine $\Delta_{(3)}^\pi$, the required
kinematic conditions are, of course, somewhat changed. The figure of
merit must still be kept high ({\em i.e.} the statistical uncertainty in the
total measured asymmetry must be kept low). In addition, the
relative contribution of $\Delta_{(3)}^\pi$ to the asymmetry must be as
large as possible relative to the statistical uncertainty in
the total asymmetry. It must also be larger than the uncertainty in
the background term $\Delta_{(2)}^\pi$. To illustrate these considerations,
we show in Fig.~\ref{d3-1d}
$\Delta_{(3)}^\pi$ as a function of $Q^2$ for various incident
energies.  This figure clearly demonstrates that $\Delta_{(3)}^\pi$ is
enhanced at lower incident energies, as we argued from basic kinematic
coefficients in Section III. In the figure, the error bar is
constructed by finding the standard deviation of the different
results for the Adler amplitudes obtained from Table~\ref{tablei}, and
should be considered a crude lower bound on the theoretical uncertainty
in this quantity.  Fig.~\ref{ad-3d} shows $\Delta_{(3)}^\pi$ as a
function of both energy and scattering angle. Again, we observe the simple
energy-dependence, as well as the relative lack of sensitivity
to scattering angle.  Tables~\ref{tableii} and \ref{tableiii} provide
more detailed numbers for a variety of kinematics, for a single
parameterization of the Adler form factors.

In Fig. \ref{adoda-3d}, we plot $A_{(3)}/\Delta A_{\rm
stat}$, versus both incident energy, $\epsilon$, and electron
scattering angle, $\theta_{\rm lab}$. Here $A_{(3)}$ is the
contribution to the asymmetry arising only from the resonant axial
transition terms: $\Delta_{(3)}^\pi\times$ the leading coefficients
in Eq.~(\ref{alr1}).
The kinematic region shown spans roughly what might
be accessible at CEBAF.  Selected numerical  values are also collected
in Tables~\ref{tableii} and \ref{tableiii}.  In order to extract
$\Delta_{(3)}^\pi$ with the greatest precision, at least three criteria
must be satisfied: (a) $A_{(3)}/\Delta A_{\rm stat}$ must be as large
as possible; (b) systematic uncertainties must be controllable at the
same level as $\Delta A_{\rm stat}/A_{\rm tot}$;
(c) the contribution from $\Delta_{(3)}^\pi$ must be considerably
larger than that of $\Delta_{(2)}^\pi$ in order to minimize the impact
of uncertainties associated with the latter. From Fig. \ref{adoda-3d}
it is clear the forward angles and moderate to high energies are favored.
Going to $\epsilon\geq 2$ GeV, however, reduces
$\Delta A_{\rm stat}/A_{\rm tot}$ to less than 1\%, a level at which
achieving similar systematic precision becomes problematic. Similarly,
the relative contributions of $\Delta_{(3)}^\pi$ and $\Delta_{(2)}^\pi$
become commensurate as $\epsilon$ increases beyond 2 GeV.


A reasonable compromise among these considerations might be the following:
incident energy in the range $1$ GeV
$< \epsilon < 2 $ GeV, scattering angle $10^o < \theta < 20^o$. The
upper bound on energy keeps $\Delta^\pi_{(3)}$ reasonably large in
comparison with $\Delta_{(2)}^\pi$,
and the smaller angles help keep the count rate, and thus the figure of
merit, high.  Throughout this kinematic range, given our
arbitrary set of experimental assumptions, $\Delta A_{\rm stat}/A_{\rm
tot}$ stays below $\approx $5\%; $A_{(3)}$ is more than a factor
of 2 larger than $\Delta A_{\rm stat}$; $A_{(3)}$ is larger than $A_{(2)}$;
and $A_{(3)}$ stays bigger than 5\% of the total
asymmetry. At the optimal kinematics points, we find
$A_{(3)}/\Delta A_{\rm stat} \approx 4$ which implies a 25\%
measurement of the axial form factor is possible. Under these conditions,
the $\Delta_{(2)}^\pi$ contribution is roughly 40\% as large as that
of $\Delta_{(3)}^\pi$, so that a 50\% uncertainty in the background
would not impair a 25\% determination of the axial response.

\section{Summary}

In this study, we have analyzed the PV $N\to\Delta$ transition with
an eye toward a prospective measurement of the PV asymmetry. In particular,
we have considered the sensitivity of $A_{LR}$ to various scenarios for
physics beyond the Standard Model -- such as leptoquarks, additional
neutral gauge bosons, and fermion compositeness -- as well as to
transition form factors of interest to hadron structure theory. After
estimating the precision with which $A_{LR}$ might be determined in
a realistic experiment, we have also estimated the scale of background
effects at the kinematics most suited for probing new electroweak physics
or hadron structure with this process.

Generally speaking, we find that, in order for $A_{LR}(N\to\Delta)$ to
compete with atomic PV or elastic PV electron scattering as a
low-energy new physics probe, a measurement at significantly better
than one percent precision would be required.  There do exist some new
physics scenarios for which $A_{LR}$ at the one percent level could
provide useful isospin information, although a measurement of the
elastic proton asymmetry would fulfill essentially the same
purpose\cite{mrm98}.
Even at the one percent level, however, a measurement would be
experimentally challenging at best.  Furthermore, it appears that
background processes are not sufficiently well understood to permit a
separation of the isovector weak charge ($\xi_V^{T=1}$) from the
non-resonant background corrections ($\Delta^\pi_{(2)}$). Future
experimental isospin decomposition of the EM pion electroproduction
process could ultimately allow a model-independent background
subtraction with sufficient precision.

The use of $A_{LR}(N\to\Delta)$ as a probe of hadron structure appears
to be a more feasible prospect at present. A $\sim 25-30\%$ determination
of the hadronic axial vector response, embedded in $\Delta^\pi_{(3)}$,
could be carried out with realistic running times. Furthermore, at
reasonable kinematics for such a measurement, {\em e.g.}
$\theta=20^\circ$, $\epsilon=1\to 2$ GeV, the backgrounds appear to
be sufficiently under control. The axial response is dominated by the
form factor ratio $C^A_5/C^V_3$, the knowledge of which would complement
information about the vector current transition form factors obtained
form EM processes. While a 25\% determination of $C^A_5/C^V_3$ would
not allow for a detailed discrimination among hadron structure model
predictions, it would significantly improve upon knowledge obtained
from charged current neutrino reactions and test model predictions at
the level of the theoretical-experimental discrepancies arising in the
vector current sector. A complete theoretical analysis of
$\Delta^\pi_{(3)}$, including the effects of potentially large and
theoretically uncertain radiative corrections associated with hadronic
PV, awaits a future study.

\acknowledgments

We thank E. Beise, J. Napolitano, and S.P. Wells for helpful
discussions. NCM and SJP thank the INT for its hospitality during the
program \lq\lq Physics Beyond the Standard Model", at which part of the
collaboration for this paper was carried out.  MJR-M wishes to thank J.
Rosner for helpful discussions regarding additional $Z$ bosons and for
the use of his oblique parameter code, and to thank W.J. Marciano for
discussions regarding the oblique parameters. The research of NCM and
JL is  supported by the U.S.  Department of Energy.  SJP was supported
in part under U.S. Department of Energy contract \#DE-FG03-93ER40774
and under a Sloan Foundation Fellowship.  MJR-M was supported in part
under U.S. Department of Energy contract \#DE-FG06-90ER40561 and under
a National Science Foundation Young Investigator Award.  HWH has been
supported by the German Academic Exchange Service
(Doktorandenstipendien HSP III). NCM was supported in part
under U.S. Department of Energy contract \#DE-FG02-88ER40448.

\appendix

\section*{Parameterization of total $\pi$ production cross section}

Although Eq.~(\ref{sigtot}) (combined with model parameters and $Q^2$
dependences for the transition form factors) provides a consistent
estimate of the total cross section, and hence the event rate for given
experimental conditions, we use a presumably more accurate estimate
based on a direct parameterization of the cross section data from
reference \cite{31}. The cross section is given by \cite{31,29},
\begin{equation}
\label{70}\frac{d^2\sigma }{d\Omega dE^{\prime }}=\Gamma _t\Sigma
\end{equation}
where
\begin{equation}
\label{71}\Gamma _t=\frac \alpha {2\pi ^2}\frac{K_0}{Q^2}\frac{E^{\prime }}{
E\left( 1-\epsilon \right) }.
\end{equation}
$K_0$ is the photon energy in the lab at $Q^2=0$:
\begin{equation}
\label{72}K_0=\frac{W^2-M^2}{2M},
\end{equation}
and we have changed notation slightly: $E$ and $E'$ are now incident and
outgoing electron energies, while $\epsilon$ is now the ``virtual photon
polarization", or
\begin{equation}
\epsilon =\frac 1{1+2\left( 1+\nu ^2/Q^2\right)
\tan^2\left( \theta /2\right) }.
\end{equation}
$\Sigma $ is
\begin{equation}
\Sigma =\sigma _T\left( q^2,W\right) +
		\epsilon \sigma _L\left( q^2,W\right),
\end{equation}
$\sigma _T$ and $\sigma _L$ are the transverse and longitudinal
cross-sections for initial photon-hadron reaction, $\nu =E-E^{\prime }$.
For the virtual photon cross-section, we use the parameterization by Brasse
{\em et al.} \cite{31}:
\begin{equation}
\label{appeq}
\Sigma =G_D^2\left( Q^2\right)
\exp\left(
        a
	+b\;\log\frac{\left| {\bf q}\right|}{K_0 }
    	+c\left( \log\frac{\left| {\bf q}\right| }{K_0}
\right) ^d
   \right) ,
\end{equation}
where the $G_D$ is the dipole form factor
\begin{equation}
\label{79}G_D\left( Q^2\right) =\left[ 1+Q^2/0.71GeV^2\right] ^{-2},
\end{equation}
and we use the following sets of parameters,  all for $W=1.23$ GeV,
depending on what the value of
$\epsilon$ is:
\begin{equation}
\begin{array}{ccccc}
\epsilon &  a & b & c & d \\
\epsilon < 0.6 & 6.149 & 1.929  & -0.087 & 3 \\
0.6 < \epsilon < 0.9 & 6.117 & 1.1866 & -0.071 & 3\\
\epsilon>0.9 & 6.125 & 1.887 & -.065 & 3
\end{array}
\end{equation}

\begin{table}
$$
\begin{array}{|c|cccc|cccc|} \hline
\vphantom{\biggl(}
& C_3^V & C_4^V & C_5^V & C_6^V & C_3^A & C_4^A & C_5^A & C_{6,non-pole}^A
\\  \hline
\text{Salin \cite{20}} & 2.0 & 0 & 0 & 0 & 0 & -2.7 & 0 & - \\
\text{Adler \cite{9,18}} & 1.85 & -0.89 & 0 & 0 & 0 & -0.3 & 1.2 & - \\
\text{Bijtebiar\cite{21}} & 2.0 & 0 & 0 & 0 & 0 & -2.9\sim -3.6 & 1.2 & - \\
\text{Zucker\cite{13}} & - & - & - & 0 & 1.8 & -1.8 & 1.9 & - \\
\text{Nath {\em et al.}\cite{22}} & 1.85 & -0.89 & 0 & 0 & 0 & -0.35 & 1.2 &
- \\
\text{Ravndal\cite{23}} & 1.70 & -1.30 & 0 & 0 & 0 & -0.65 & 0.97 & - \\
\text{Orsay {\em et al.}\cite{24}} & 1.54 & -1.17 & 0 & 0 & 0 & -0.20 & 0.83
& -
\\
\text{K\"orner {\em et al.}\cite{25}}
& 1.70 & -1.30 & 0 & 0 & 0 & -0.32 & 0.97 & - \\
\text{Jones and Petcov\cite{5}} & 2.05 & -1.56 & 0 & 0 & 0 & -0.3 & 1.2 & -
\\
\text{HHM\cite{HHM}} & 1.39 & -1.06 & 0 & 0 & 0 & -0.29\pm 0.006
& 0.87\pm 0.03 & - \\
\text{SU(6)}\cite{paper1} & 1.48 & -1.13 & 0 & 0 & 0 & -0.38 & 1.17 & - \\
\text{IK}\cite{paper1}
& 1.32 & -0.79 & -0.36 & 0.014 & -0.0013 & -0.66 & 1.16 & 0.032 \\
\text{IK2}\cite{paper1}
& 1.37 & -0.66 & -0.59 & -0.015 & 0.0008 & -0.657 & 1.20 & 0.042
\\
\text{D-mixing}\cite{paper1} & 1.29
& 0.78 & -1.9 & -0.15 & 0.052 & 0.052 & 0.813 & -0.17 \\ \hline
\end{array}
$$
\caption{Values of Adler form factors at $Q^2=0$
in various empirical and theoretical
approaches. \lq\lq $-$" means the reference does not provide a prediction.
As
discussed in the text, $C_{6}^A$ is a pseudoscalar response, and effectively
does not contribute to the PV asymmetry. \lq\lq Non-pole" indicates that
the pion pole contributions to $C_{6}^A$ have not been included in the
predictions cited (see Ref.~\protect\cite{Abd72}).}
\label{tablei}
\end{table}

\begin{table}
$$
\begin{array}{|ccc|ccccc|} \hline
\vphantom{\Biggl( }
E\left( {\rm GeV}\right) &
\theta_{\rm lab}(^o) &
Q^2 ({\rm GeV}^2) &
10^5 A_{\rm tot} &
\frac{\delta A_{\rm stat}}{A_{tot}}(\%) &
\frac{\Delta^\pi_{(2)}}{\Delta^\pi}(\%) &
\frac{ \Delta_{(3)}^\pi}{\Delta^\pi}(\%) &
\frac{ A_{(3)}}{\delta A_{stat}} \\
\hline
  .4 &  10. &  .001 &   -.01 & 148.5 & -.19  &23.2 &   .2 \\
  .5 &  10. &  .002 &   -.03 &  45.9 & -.24  &20.4 &   .4 \\
  .6 &  10. &  .005 &   -.06 &  24.2 & -.30  &17.7 &   .7 \\
  .7 &  10. &  .008 &   -.09 &  15.1 & -.37  &15.5 &  1.0 \\
  .9 &  10. &  .015 &   -.17 &   7.7 & -.55  &12.2 &  1.6 \\
\hline
  .4 &  15. &  .002 &   -.02 & 100.1 & -.22  &23.2 &   .2 \\
  .5 &  15. &  .005 &   -.07 &  31.2 & -.33  &20.3 &   .7 \\
  .6 &  16. &  .012 &   -.14 &  15.7 & -.46  &17.6 &  1.1 \\
  .7 &  16. &  .019 &   -.22 &   9.9 & -.61  &15.3 &  1.5 \\
  .9 &  15. &  .033 &   -.37 &   5.5 & -.98  &12.1 &  2.2 \\
\hline
  .4 &  20. &  .003 &   -.04 &  76.3 & -.27  &23.1 &   .3 \\
  .5 &  20. &  .009 &   -.12 &  24.0 & -.45  &20.2 &   .8 \\
  .6 &  20. &  .018 &   -.22 &  13.0 & -.67  &17.5 &  1.3 \\
  .7 &  20. &  .029 &   -.34 &   8.3 & -.92  &15.2 &  1.8 \\
  .9 &  20. &  .058 &   -.64 &   4.5 & -1.50  &12.0 &  2.7 \\
\hline
  .4 &  45. &  .013 &   -.16 &  38.9 & -.61  &22.4 &   .6 \\
  .5 &  45. &  .041 &   -.49 &  13.2 & -1.34 &19.1 &  1.5 \\
  .6 &  45. &  .078 &   -.90 &   7.6 & -2.06 &16.4 &  2.1 \\
  .7 &  45. &  .122 &  -1.38 &   5.4 & -2.76 &14.2 &  2.6 \\
  .9 &  45. &  .232 &  -2.53 &   3.8 & -4.04  &11.2 &  2.9 \\
\hline
  .4 &  90. &  .035 &   -.43 &  28.0 & -1.32  &21.0 &   .8 \\
  .5 &  90. &  .106 &  -1.24 &  10.9 & -2.81 &17.1 &  1.6 \\
  .6 &  90. &  .192 &  -2.18 &   7.4 & -3.92 &14.3 &  1.9 \\
  .7 &  90. &  .291 &  -3.21 &   6.0 & -4.76 &12.2 &  2.0 \\
  .9 &  90. &  .517 &  -5.53 &   5.2 & -5.95 & 9.3 &  1.8 \\
 \hline
  .4 & 180. &  .054 &   -.65 &  26.7 & -1.88 &19.8 &   .7 \\
  .5 & 180. &  .157 &  -1.81 &  11.6 & -3.70  &15.5 &  1.3 \\
  .6 & 180. &  .276 &  -3.08 &   8.5 & -4.82 &12.7 &  1.5 \\
  .7 & 180. &  .407 &  -4.43 &   7.5 & -5.58 &10.7 &  1.4 \\
  .9 & 180. &  .694 &  -7.33 &   7.3 & -6.58 & 8.1 &  1.1 \\
\hline
\end{array}
$$
\caption{Columns 4-8 give SM prediction of the asymmetry,
the experimental statistical uncertainty for the asymmetry (given the
assumptions of section VIII),
the percentage contribution of the vector current backgrounds,
the percentage contribution of the axial-vector excitation of the Delta,
and ratio of axial contribution to statistical uncertainty, respectively,
as functions of energy and scattering angle.
$Q^2$ is calculated assuming we are sitting on the $\Delta$ peak.
$\Delta^\pi=\Delta^\pi_{(1)}+\Delta^\pi_{(2)}+\Delta^\pi_{(3)}$, and
$A_{(3)}$ is the contribution to the asymmetry arising from
$\Delta^\pi_{(3)}$.
The axial contribution uses the parameter set of
Adler-Kitagaki, but assumes $C_4^V$ is constrained
by the quark model relation, Eq.~(\protect\ref{57}).}
\label{tableii}
\end{table}

\begin{table}
$$
\begin{array}{|ccc|ccccc|} \hline
\vphantom{\Biggl( }
E\left( {\rm GeV}\right) &
\theta_{\rm lab}(^o) &
Q^2 ({\rm GeV}^2) &
10^5 A_{\rm tot} &
\frac{\delta A_{\rm stat}}{A_{tot}}(\%) &
\frac{\Delta^\pi_{(2)}}{\Delta^\pi}(\%) &
\frac{ \Delta_{(3)}^\pi}{\Delta^\pi}(\%) &
\frac{ A_{(3)}}{\delta A_{stat}} \\
\hline
 1.0 &  10. &  .020 &   -.22 &   5.9 & -.7   &11.0 &  1.9 \\
 2.0 &  10. &  .098 &  -1.01 &   1.4 & -2.2  & 5.5 &  3.8 \\
 3.0 &  10. &  .231 &  -2.33 &    .8 & -4.0  & 3.6 &  4.3 \\
 4.0 &  10. &  .418 &  -4.17 &    .7 & -5.5  & 2.6 &  4.0 \\
 \hline
 1.0 &  15. &  .044 &   -.48 &   4.3 & -1.2  &10.9 &  2.5 \\
 2.0 &  15. &  .211 &  -2.17 &   1.3 & -3.8  & 5.4 &  4.1 \\
 3.0 &  15. &  .491 &  -4.94 &   1.0 & -5.9  & 3.5 &  3.6 \\
 4.0 &  15. &  .872 &  -8.68 &   1.0 & -7.4  & 2.4 &  2.5 \\
\hline
 1.0 &  20. &  .075 &   -.82 &   3.6 & -1.8  &10.9 &  3.0 \\
 2.0 &  20. &  .355 &  -3.64 &   1.4 &  -5.0 & 5.3 &  3.8 \\
 3.0 &  20. &  .808 &  -8.11 &   1.3 & -7.1  & 3.3 &  2.5 \\
 4.0 &  20. & 1.406 & -13.96 &   1.6 & -8.9  & 2.2 &  1.4 \\
\hline
 1.0 &  45. &  .296 &  -3.19 &   3.4 & -4.6  &10.0 &  2.9 \\
 2.0 &  45. & 1.199 & -12.18 &   3.5 & -8.0  & 4.4 &  1.3 \\
 3.0 &  45. & 2.416 & -24.02 &   5.3 & -12.5 & 2.3 &   .4 \\
 4.0 &  45. & 3.816 & -37.54 &   8.7 & -36.6  & 1.3 &   .2 \\
\hline
 1.0 &  90. &  .641 &  -6.79 &   5.2 & -6.4  & 8.3 &  1.6 \\
 2.0 &  90. & 2.123 & -21.30 &   8.9 & -11.2  & 3.2 &   .4 \\
 3.0 &  90. & 3.805 & -37.52 &  16.1 & -32.7  & 1.5 &   .1 \\
 4.0 &  90. & 5.566 & -54.46 &  26.8 & -13.4  &  .8 &   .0 \\
\hline
 1.0 & 180. &  .846 &  -8.85 &   7.5 & -7.0  & 7.1 &   .9 \\
 2.0 & 180. & 2.526 & -25.21 &  14.5 & -11.9  & 2.7 &   .2 \\
 3.0 & 180. & 4.320 & -42.48 &  26.5 & -33.9  & 1.3 &   .0 \\
 4.0 & 180. & 6.150 & -60.08 &  43.8 & -10.4  &  .6 &   .0 \\
\hline
\end{array}
$$
\caption{Same as previous table, for higher incident energies.}
\label{tableiii}
\end{table}

\widetext
\begin{figure}[htb]

\vspace{1.0cm}

\begin{center}
\epsfysize=6cm
~\epsffile{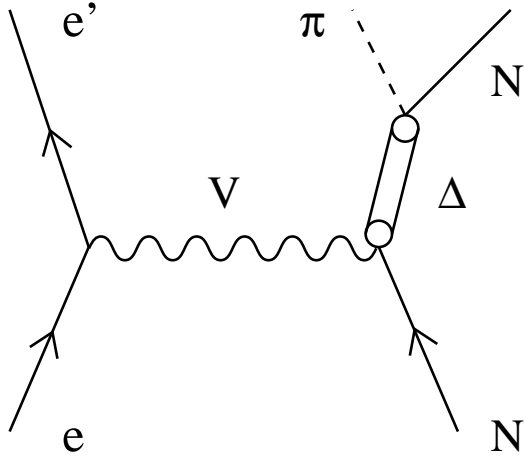}
\end{center}
\caption{Feynman diagram for tree-level electroweak
excitation of the $\Delta$ resonance. }
\label{fig1}
\end{figure}

\widetext
\begin{figure}[htb]
\begin{center}
\epsfysize=12cm
~\rotate[r]{\epsffile{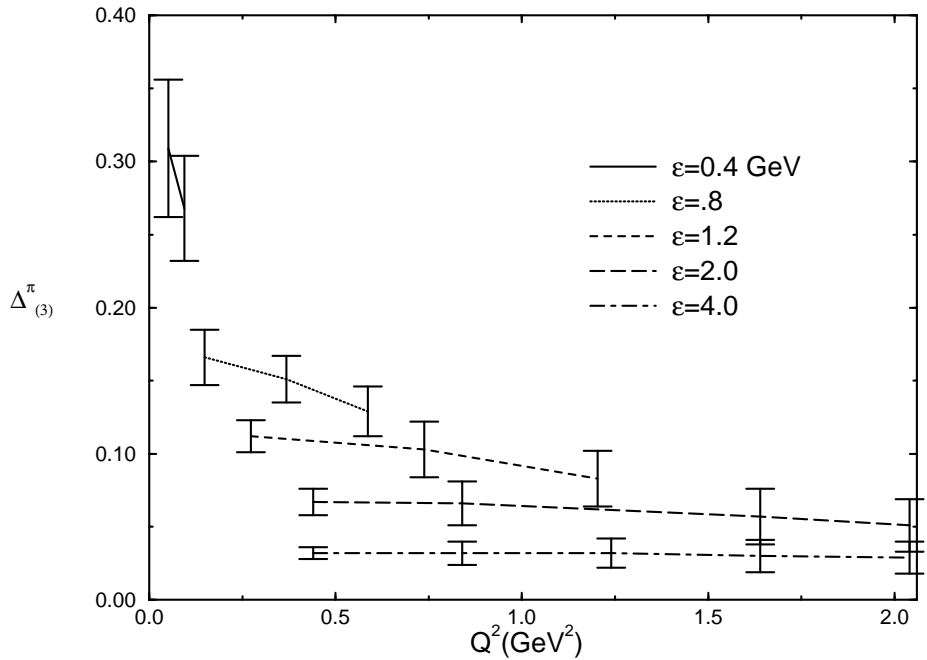}}
\end{center}

\vspace{0.5cm}

\caption{ $\Delta_{(3)}^\pi$ as a function of $Q^2$ in for
different incident electron energies, $\epsilon$.
We computed transition form factors using the models of Table I to
generate a rough spread of
theoretical results, shown by the error bars.}
\label{d3-1d}
\end{figure}

\widetext
\begin{figure}[htb]

\vspace{-5cm}

\begin{center}
\epsfysize=8cm
~\epsffile{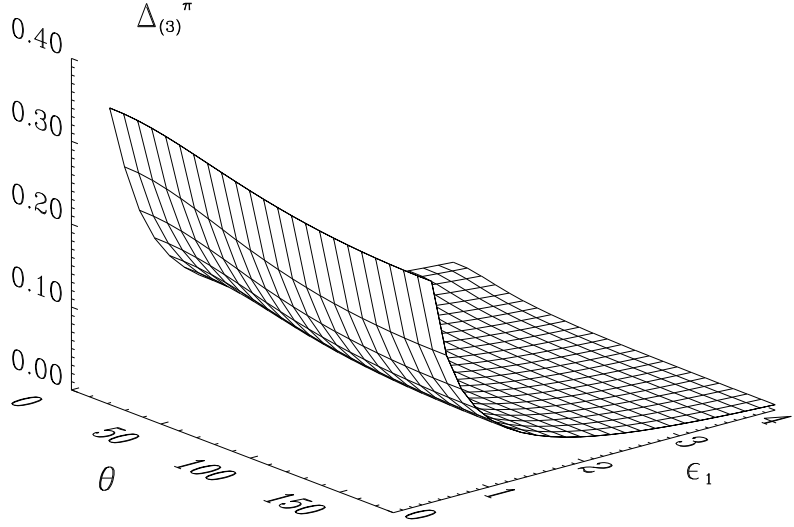}
\end{center}
\caption{ As in previous figure, with
$\Delta_{(3)}^\pi$ plotted in 3 dimensions
versus both incident energy, $\epsilon$, and electron scattering
angle, $\theta_{\rm lab}$. We have computed $\Delta_{(3)}^\pi$ only with the
Adler-Kitagaki \protect\cite{18} parametrization here.}
\label{ad-3d}
\end{figure}

\widetext
\begin{figure}[htb]

\vspace{-2.5cm}

\begin{center}
\epsfysize=9.5cm
\rotate[r]{\epsffile{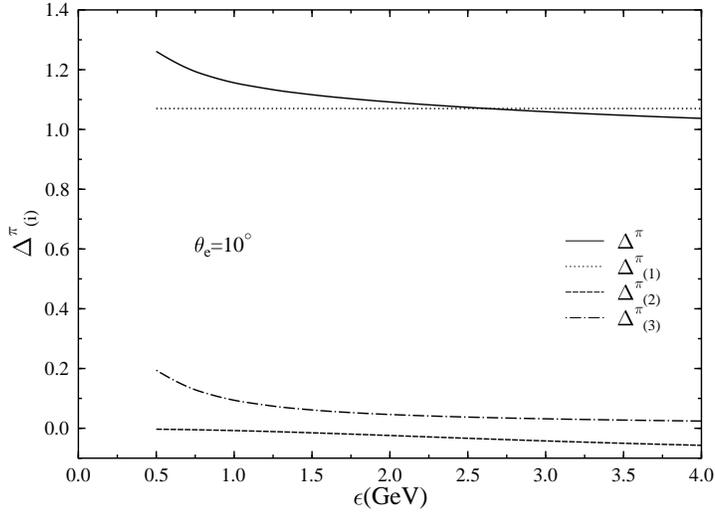}}\quad
\end{center}

\vspace{-4.2cm}

\caption{Contributions to the asymmetry from the $\Delta^\pi_{(i)}$
defined in the text as a function of incident energy for $\theta
=10^\circ$. Here, $\Delta^\pi=
\Delta^\pi_{(1)}+\Delta^\pi_{(2)}+\Delta^\pi_{(3)}$.}
\end{figure}

\newpage

\begin{figure}[htb]

\vspace{1.0cm}

\begin{center}
\epsfysize=7cm
~\epsffile{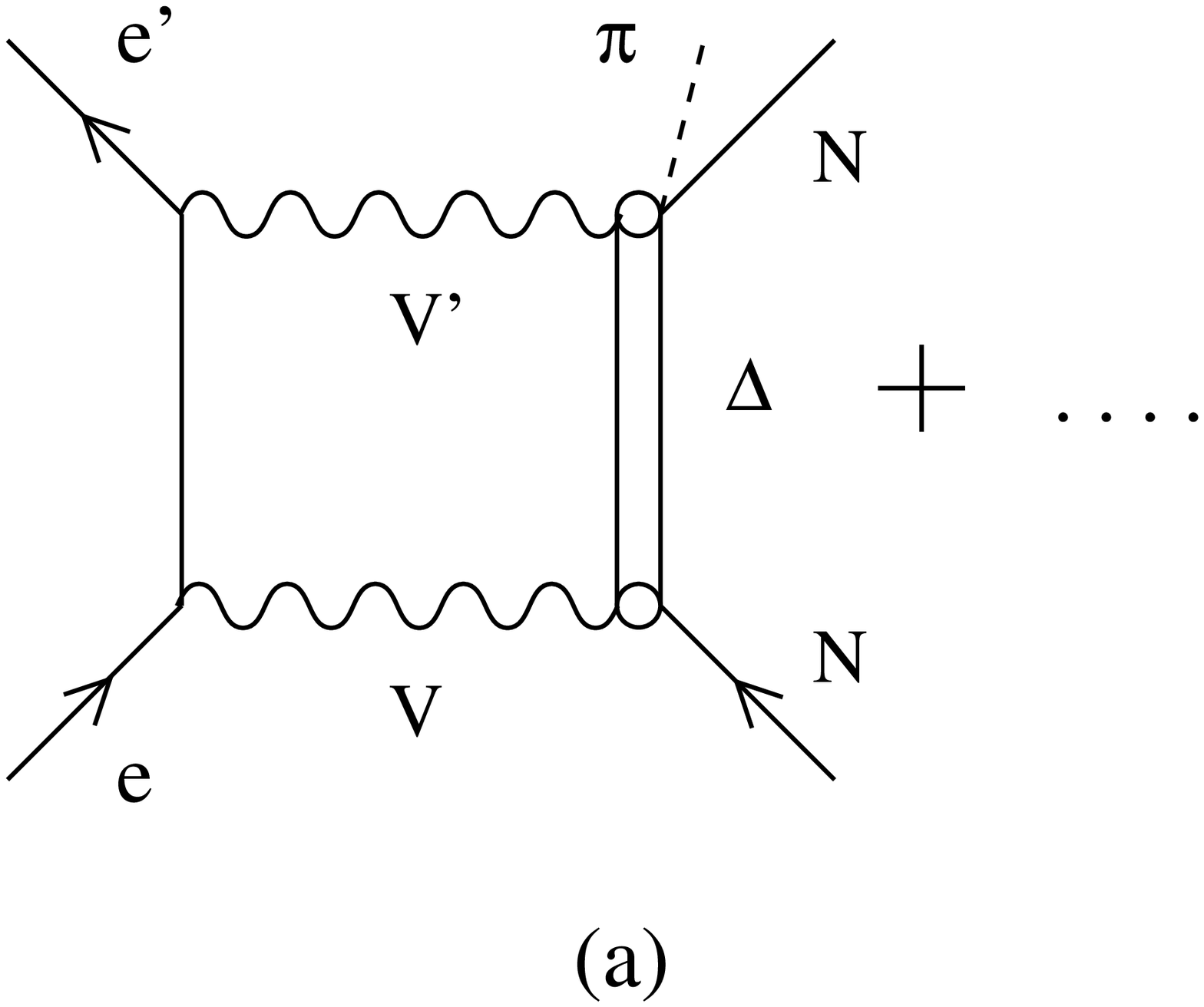}
\end{center}

\vspace{0.3cm}

\begin{center}
\epsfysize=6cm
~\epsffile{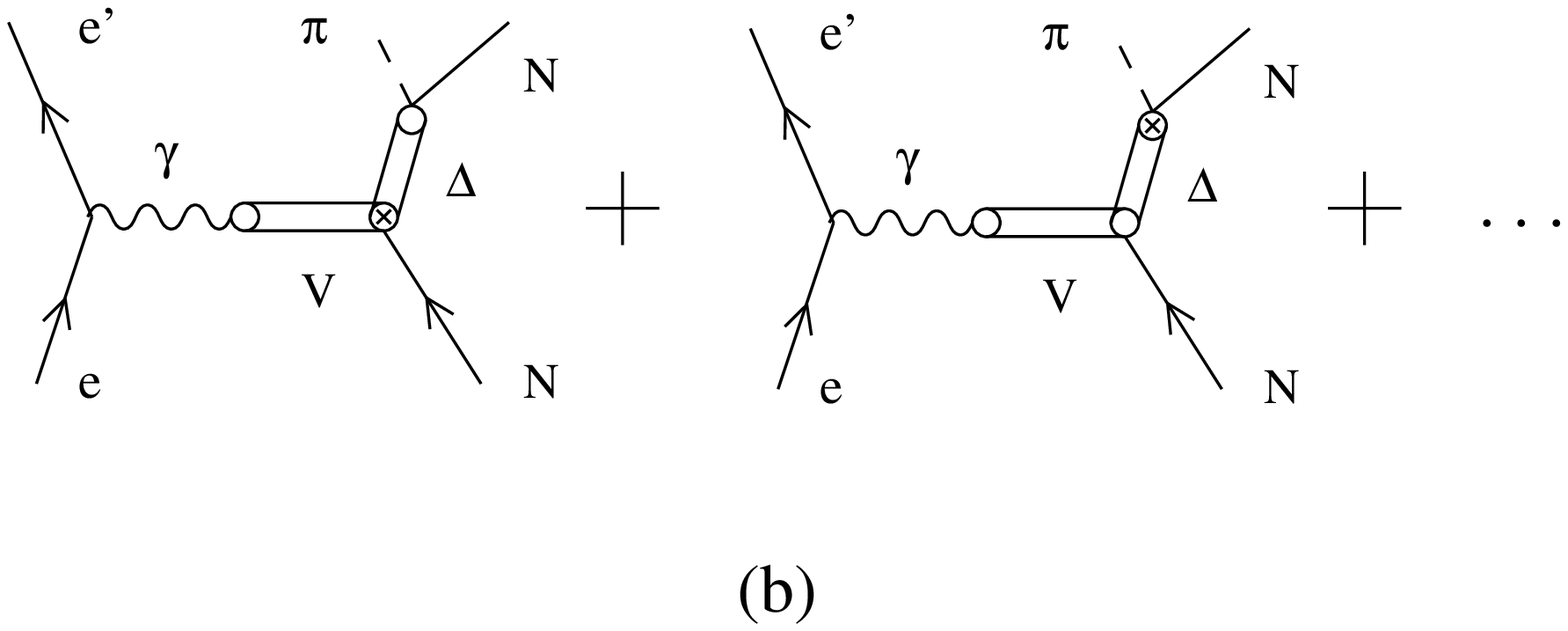}
\end{center}

\caption{Higher-order, hadron structure-dependent electroweak corrections
to the asymmetry. Fig (a) shows two-boson exchange \lq\lq dispersion
correction", where $V,V'=\gamma, Z^0, W^\pm$. Fig (b) illustrates corrections
arising from hadronic parity violation. Here, open circle indicates a
parity-conserving coupling and $\otimes$ denotes a weak, PV coupling.
Here, $V$ indicates a vector meson such as the $\rho$ or $\omega$.}
\end{figure}

\widetext
\begin{figure}[htb]
\begin{center}
\epsfysize=9cm
~\epsffile{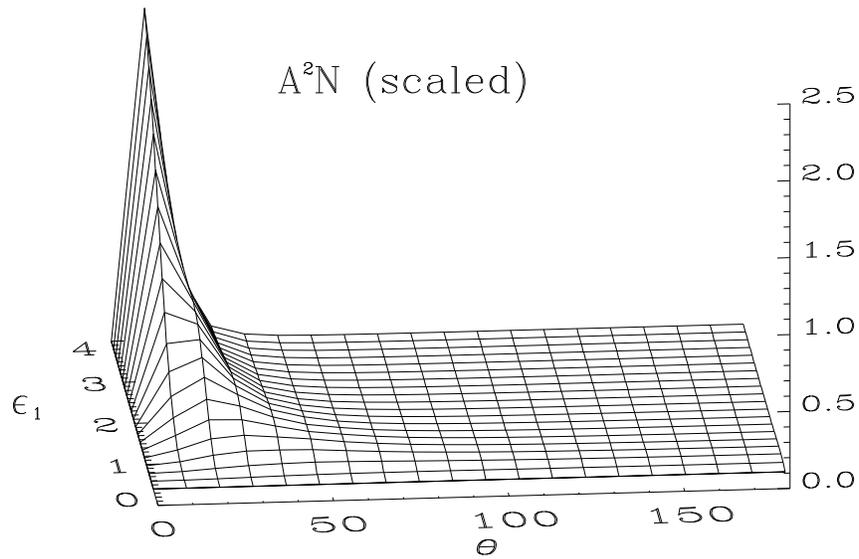}
\end{center}

\vspace{-0.5cm}

\caption{ As in Figure \protect{\ref{ad-3d}}, a 3-D plot of
a measure of the figure of merit,
$A^2 N$, scaled by
$10^4$, versus incident energy and electron
scattering angle.}
\label{fom-3d}
\end{figure}

\widetext
\begin{figure}[htb]

\vspace{-1.0cm}

\begin{center}
\epsfysize=9cm
~\epsffile{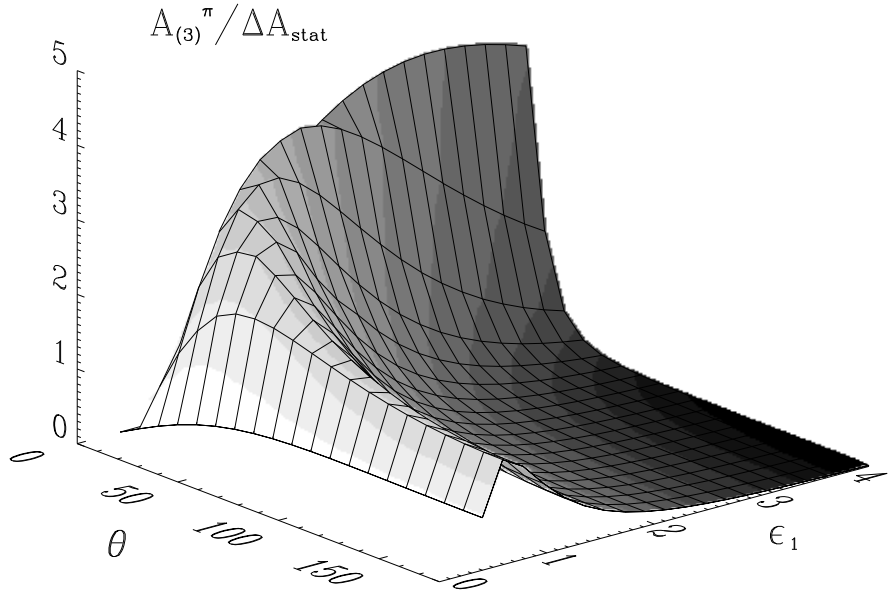}
\end{center}
\caption{ As above, a 3-D plot of
$A_{(3)}/A_{\rm tot}\over \delta A_{\rm stat}/A_{\rm tot}$,
versus both incident energy and electron scattering
angle. (The shading is determined by the value of $\Delta_{(3)}^\pi$,
smaller values are shaded darker) }
\label{adoda-3d}
\end{figure}

\end{document}